\documentclass[onecolumn,showpacs,floatfix,10pt,nofootinbib]{revtex4}
\usepackage{amssymb,amsmath}
\usepackage[dvips]{graphics,color}
\usepackage[colorlinks,hyperindex]{hyperref}
\usepackage{epsfig}\usepackage{float,stmaryrd,textcomp,bm,mathbbol}\usepackage{float,upgreek,yfonts}
\usepackage{graphicx}
\newcommand{\f}{\frac}
\newcommand{\ba}{\begin{eqnarray}}
\newcommand{\ea}{\end{eqnarray}}
\newcommand{\bege}{\begin{equation}}
\newcommand{\enge}{\end{equation}}

\newcommand{\beq}{\begin{eqnarray}}
\newcommand{\benu}{\begin{enumerate}}
\newcommand{\enu}{\end{enumerate}}
\newcommand{\eeq}{\end{eqnarray}}

\newcommand{\br}{\langle\,0\,|}
\newcommand{\ke}{|\,0\,\rangle}

\hypersetup{hidelinks,backref=true,pagebackref=true,hyperindex=true,colorlinks=true,breaklinks=true,urlcolor= blue}
\hypersetup{%
  colorlinks = true,
  linkcolor  = blue
}

\newcommand{\RR}{\mathbb{R}}

 \newcommand{\bea}{\begin{eqnarray}}
 \newcommand{\eea}{\end{eqnarray}}
 \newcommand{\pwp}{\uppsi}
 \newcommand{\wpwp}{\bar\uppsi} 
 \newcommand{\pwpw}{\uppsi_1}
 \newcommand{\pwpe}{\uppsi_2}
\usepackage{hyperref}
\usepackage{xcolor}

 \def\bra#1{\left\langle#1\right|}

\begin{document}

\title{Opening the Pandora's box of quantum spinor fields}

\author{L. Bonora}
\email{bonora@sissa.it}
\affiliation{International School for Advanced Studies - SISSA, Via Bonomea 265,
34136 Trieste, Italy.}
\author{J. M. Hoff da Silva}
\email{hoff@feg.unesp.br} 
\affiliation{Departamento de F\'{\i}sica e Qu\'{\i}mica, Universidade Estadual Paulista - UNESP, Guaratinguet\'{a}, SP, Brazil.}
\author{R. da Rocha}
\email{roldao.rocha@ufabc.edu.br}
\affiliation{Centro de Matem\'atica, Computa\c c\~ao e Cogni\c c\~ao, Universidade Federal do ABC - UFABC,  09210-580, Santo Andr\'e, Brazil.}


\begin{abstract}
 Lounesto's classification of spinors \textcolor{black}{is}  a comprehensive and exhaustive  \textcolor{black}{algorithm that, based on the bilinears covariants, discloses the possibility of a large variety of spinors, comprising} regular and singular spinors and their unexpected applications in physics and including the cases of Dirac, Weyl, and Majorana as very particular spinor fields. In this paper we pose \textcolor{black}{the problem of an analogous classification in the framework of second quantization.} \textcolor{black}{We first discuss in general the nature of the problem. Then we start the analysis of two basic bilinear covariants, the scalar and pseudoscalar,}  in the second quantized setup, with expressions applicable to the quantum field theory extended to all types of spinors. \textcolor{black}{One can see that an ampler set of possibilities opens up with respect to the classical case.} A quantum reconstruction algorithm is also proposed. The Feynman \textcolor{black}{propagator is} extended for spinors in all classes.
\end{abstract}
\maketitle

\section{Introduction}

Classical spinors can be classified on the basis of the bilinear covariants constructed out of them, satisfying the Fierz identities. This property has led to the well-known Lounesto's classification into six classes \cite{lou2}. While the possibilities raised by these classes are still being explored, a natural question arises: does quantization preserve this classification, or does it open new possibilities? In this paper we would like to start exploring this problem. A preliminary question is of course how to formulate the classification problem in a quantum context.  In classical or first quantized theory, once we know the bilinears,  the reconstruction theorem guarantees that one can explicitly construct (up to a phase) spinors for each class \cite{Taka,Cra}. Therefore knowing the spinor bilinears is equivalent to knowing the spinor itself, (up to a phase) there is nothing else to be known from the point of view of the observables.  In the case of second quantized spinors, new features may appear. In fact, in quantum theory, knowing the theory means knowing all the correlators. Therefore a classification of quantum spinors must be based on the knowledge 
of their correlators. A possible question seems therefore to be  whether the set of correlators of the bilinears provides a basis for a classification.   This sounds sensible, but seems to be beyond our reach, for the moment.

A more viable approach seems to be the perturbative one. From a perturbative point of view one of the basic ingredients of a quantum field theory are the propagators. In free field theories, knowing the propagators one can compute all the correlators via the Wick theorem. In interacting theories
propagators are also the basic ingredients together with vertices in order to evaluate correlators. Moreover a propagator contains the information about the equation of motion, being the inverse of the kinetic operator. Thus it is sensible to focus on the (free) fermion propagators. In order to evaluate a free propagator we need an expansion of the spinor field in plane waves, which in turn requires that the spinor satisfies the Klein-Gordon (KG) equation. This is our basic requirement, that the spinors satisfy the KG equation (not necessarily the Dirac or other linear equations). Classifying the propagators requires classifying such (quantum) spinors. In the sections that follow, to start with, we will construct the pseudoscalar and the scalar bilinear covariants, constructed out of such arbitrary second quantized free spinor fields at different points, and analyze them. In this way we come across the bilinears of the expansion coefficients, which are ordinary classical spinors to which the Lounesto classification can be applied. This leads to a new puzzling aspect: the expansion coefficients are classical spinors, but they  depend on the momentum, not on the coordinates. Since the Lounesto classification is entirely algebraic (it does not depend on the spacetime coordinates), it holds independently of spacetime and momentum space. However the two classifications are unrelated. This aspect of our approach needs to be further investigated.  Anyway, assuming the usual plane wave expansion where the spinors, used as expansion coefficients, can run throughout all the regular and singular spinors in the first quantized formalism, we show that the second quantized regular and singular spinors are split into additional  subclasses of fields. This is the main result of our paper. Needless to say one should analyze also the other bilinears, not only the scalar and pseudoscalar. But the latter are enough to appreciate the wealth of possibilities that open up when the classification algorithm is applied to second quantized spinors.

In classical theory, once we know the bilinears, which is equivalent to know the spinor itself, there is nothing else to be known from the point of view of the observables. In the case of second quantized spinors, new aspects should be introduced, including anti-commutativity of the creation and annihilation operators and, eventually, correlators of bilinears. In order to classify quantum spinors, one should know their correlators. For free theories, knowing the propagator one can compute all the correlators via the Wick theorem, but the propagator contains the information about the equation of motion, being the inverse of the kinetic operator. 
One of the outputs here presented is an endeavor to formulate quite general fermionic interacting terms to Lagrangians describing quantum processes. In fact, the possibilities raised in the scope of Lounesto's classification deserve to be addressed in a quantum theory. We focus in the (indeed possible) plurality of the expansion coefficients to analyse the scalar and pseudoscalar spinorial bilinear covariants obtained from the quantum operators. Thus, by working out the underlying algebra and constraints to be respected by the spinorial coefficients, we are able to explore the plethora of quantum possibilities for couplings built upon these quantities.    

This paper is organized as follows: in Sect. II a brief review of the Lounesto's classification of spinors is presented, with special attention to the flagpole, flag-dipole and dipole structures of singular spinors, the Fierz aggregate and the reconstruction theorem. Sect. III contains a general discussion of how to formulate a classification of quantum spinors and how this naturally leads to the problem of classifying the spinors in momentum space. Sect. IV is devoted to classify the quantum fields into singular and regular second quantized fields, where the reconstruction theorem plays a central role in the refinement of the analysis. Assuming the usual plane wave expansion, typical for a free field analysis, whose spinors used as expansion coefficients can run throughout all the regular and singular spinors in the first quantized formalism, we show that the second quantized regular and singular spinors are split into more subclasses of fields. In Sect. V we extend the calculations of $n$-point functions and propagators to all the spinors in the Lounesto's classification that satisfy the Dirac equation, as well as for eigenspinors of the charge conjugation operator that have mass dimension 3/2 in the four dimensional Minkowski spacetime. In Sect. VI explicit expressions for the normal ordered  bilinear covariants of arbitrary quantum fields are obtained, paving the way for a second quantized version of the reconstruction theorem. In Sect. VII the conclusions are discussed, recalling the main important results throughout the previous sections, and perspectives are outlined.

\section{General Bilinear Covariants and Spinor Field Classes}

Let $(M,\eta)$ be a (oriented) manifold, with   tangent bundle $TM$ and a metric $\eta: \sec TM\times \sec TM\to\RR$, admitting an exterior bundle 
$\Omega(M)$ with sections $\sec\Omega(M)$. The Clifford product, between an arbitrary 1-form $v \in 
\sec\Omega^1(M)$ and an arbitrary form  $\zeta \in \sec\Omega(M)$, can be expressed by the exterior product and the contraction, namely, $v \zeta = v \wedge \zeta+ v 
\lrcorner\, \zeta$ and $\zeta v=\zeta\wedge v + \zeta\llcorner v$, where $\eta(\zeta_1\lrcorner\zeta_2,\zeta_3)=\eta(\zeta_1,\zeta_2\wedge\zeta_3)$, for all $\zeta_1, \zeta_2, \zeta_3\in \sec\Omega(M)$. Considering the Minkowski spacetime $M$, the set  $\{e^{\mu }\}$ is hereon a basis for the section of the coframe bundle
${P}_{\mathrm{SO}_{1,3}^{e}}(M)$. Classical \textcolor{black}{Dirac} spinor fields  carry the 
$\rho = {\left(\frac12,0\right)}\oplus {\left(0,\frac12\right)}$ representation of the Lorentz group. For arbitrary  
spinor fields  
$\psi \in \sec {P}_{\mathrm{Spin}_{1,3}^{e}}(M)\times_{\rho}
\mathbb{C}^{4}$, the bilinear covariants, defined at each point $x\in M$, read
\begin{subequations}
\begin{eqnarray}
\sigma(x) &=& \bar{\psi}(x)\psi(x)\,,\label{sigma}\\
J_{\mu }(x) e^{\mu }=\mathbf{J}(x)&=&\bar{\psi}(x)\gamma _{\mu }\psi(x)\, e
^{\mu}\,,\label{J}\\
S_{\mu \nu }(x) e^{\mu}\wedge e^{ \nu }=\mathbf{S}(x)&=&\tfrac{1}{2}i\bar{\psi}(x)\gamma
_{\mu
\nu }\psi(x) \, e^{\mu }\wedge  e^{\nu }\,,\label{S}\\
 K_{\mu}(x) e^{\mu }=\mathbf{K}(x)&=&i\bar{\psi}(x)\gamma_{0123}\gamma _{\mu }\psi(x)
\, e^{\mu }\,,\label{K}\\\omega(x)&=&-\bar{\psi}(x)\gamma_{0123}\psi(x)\,,  \label{fierz}
\end{eqnarray}\end{subequations}
where $\bar\psi=\psi^\dagger\gamma_0$, $\gamma
_{\mu
\nu }=\frac{i}{2}[\gamma_\mu, \gamma_\nu]$, 
$\gamma_{0123}:=\gamma_0\gamma_1\gamma_2\gamma_3$ and $\gamma_{\mu }\gamma _{\nu
}+\gamma _{\nu }\gamma_{\mu }=2\eta_{\mu \nu }\mathbf{1}$. 
Regarding the electron theory,  ${\bf J}$ is a conserved current, due to the U(1) symmetry of the Dirac theory. Hence, $J_0 = \psi^\dagger\psi$ provides
the probability density associated with the electron, which should not vanish. It is worth emphasizing that the reason for  considering ${\bf J}$  as the  current density is clear in the case when the spinor obeys the Dirac equation \cite{Villalobos:2015xca}. The mass dimension  $3/2$ in this case is the usual one to spin-1/2 fermions, in the standard model elementary particles. When ${\bf J}=0$ is required, the underlying dynamics is not governed by the  Dirac equation, being also implicit that 
this case precludes mass dimension 3/2 spinors, being restricted to mass dimension one spinors \cite{Villalobos:2015xca}. Since the construction is relativistic, the eventual emergent spinors  with ${\mathbf J}=0$ are anyway ruled by the Klein--Gordon equation.

  The Fierz identities read
\begin{equation}\label{fifi}
\mathbf{K}(x)\wedge\mathbf{J}(x)=(\omega(x)-\sigma(x)\star)\mathbf{S}(x),\qquad\mathbf{J}^
{2}(x)=\omega^{2}(x)+\sigma^{2}(x),\qquad\mathbf{K}^{2}(x)+\mathbf{J}^{2}(x)
=0=\mathbf{J}(x)\cdot\mathbf{K}(x)
\,.  
\end{equation}
\noindent When either $\omega\neq0$ or $\sigma\neq0$ [$\omega=0=\sigma$] the
spinor field $\psi$ is named regular [singular] spinor and also satisfy:
\beq
\mathbf{S}(x)\llcorner\mathbf{J}(x)  & =&\omega(x)\mathbf{K}(x),\qquad\quad\qquad\qquad\quad\mathbf{S}(x)%
\llcorner\mathbf{K}(x)=\omega(x)\mathbf{J}(x),\qquad\quad\qquad\qquad\quad
\star\mathbf{S}(x)\llcorner\mathbf{J}(x)=-\sigma(x)\mathbf{K}(x),\nonumber\\\star\mathbf{S}(x)\llcorner\mathbf{K}(x)  &=&-\sigma(x)\mathbf{J}(x),\qquad\quad\qquad\qquad
\mathbf{S}(x)\llcorner\mathbf{S}(x)=-\omega^{2}(x)+\sigma^{2}(x),\qquad\qquad\qquad 
\star\mathbf{S}(x)\llcorner\mathbf{S}(x)=2\omega(x)\sigma(x)\gamma_5,\nonumber\\
\mathbf{S}(x)\mathbf{K}(x)  &=&(\omega(x)-\sigma(x)\star)\mathbf{J}(x),\qquad\qquad\mathbf{S}^{2}(x)=\omega^{2}(x)-\sigma^{2}(x)
-2\omega(x)\sigma(x)\gamma_{0123}. \label{FIERZ}%
\eeq
The bilinear covariants are physically interpreted in the Dirac theory. In fact, $eJ_0$ is the charge density, whereas $ec J_k$  is identified to the (electric) current density. The quantity $\frac{e\hbar}{2mc} S^{ij}$ is the magnetic moment density, while $\frac{e\hbar}{2mc} S^{0j}$ is the electric moment density. The $(\hbar/2) K_\mu$ is interpreted as chiral current, conserved when $m=0$. The interpretation of the scalar $\sigma$ and pseudoscalar $\omega$ bilinear covariants is less clear, but when combined into $\rho^2 = \sigma^2 + \omega^2 = |J|^2$ (by the Fierz--Pauli--Kofink (FPK) identities), $\rho$ can be interpreted as probability density for regular spinors \cite{lou2}.

Lounesto classified spinor fields into six disjoint classes, wherein  $\mathbf{J}\neq 0$ \cite{lou2,Vaz:2016qyw}:
\begin{subequations}\beq
1)&& \sigma(x)\neq0,\;\;\;\omega(x)\neq0,\;\;\;\mathbf{S}(x)\neq 0,\;\;\;\mathbf{K}(x)\neq0,\label{tipo1}\\
2)&& \sigma(x)\neq0,\;\;\;
\omega(x) = 0,\;\;\;\mathbf{S}(x)\neq 0,\;\;\;\mathbf{K}(x)\neq0,\label{tipo2}\\
3)&&\sigma(x)= 0, \;\;\;\omega(x) \neq0,\;\;\;\mathbf{S}(x)\neq 0,\;\;\;\mathbf{K}(x)\neq0,\label{tipo3}\\
4)&&\sigma(x)=0,\;\;\;\omega(x)=0,\;\;\;\mathbf{S}(x)\neq 0,\;\;\;\mathbf{K}(x)\neq0,
\label{tipo4}\\
5)&&\sigma(x)=0,\;\;\;\omega(x)=0,\;\;\;\mathbf{S}(x)\neq0,\;\;\;
\mathbf{K}(x)=0,
\label{tipo5}\\
6)&&\sigma(x)=0,\;\;\;\omega(x)=0,\;\;\; \mathbf{K}(x)\neq0,
\;\;\; \mathbf{S}(x)=0.\label{tipo6}\eeq\end{subequations}
\noindent  
Singular spinor fields in the above Lounesto's classes 4, 5, and 6, 
 are, respectively, identified to flag-dipoles, flagpoles, and
dipoles structures.  In fact, the definitions (\ref{J}) and (\ref{K})  \textcolor{black}{identify} the current density  {\bf J} and  the chiral current {\bf K} as 1-form fields, they \textcolor{black}{are referred to} as poles, \textcolor{black}{while} the 2-form field ${\bf S}$ is a flag, due to its bivector structure. Hence, type-5 spinor fields, having a null pole ${\bf K} = 0$, the pole ${\bf J}\neq0$ and ${\bf S}\neq 0$ are called flag-poles. Type-4 spinor fields have instead two poles, ${\bf J}\neq0$ and ${\bf K}\neq0$, and the flag ${\bf S}\neq0$, corresponding therefore to a flag-dipole. These objects encode Penrose flagpole structures, constructed upon pure spinors \cite{Vaz:2016qyw}. Regarding type-6 spinors, still ${\bf J}\neq0$ and  ${\bf K}\neq0$, however the flag {\bf S} is zero, terming it a dipole spinor field. Flag-dipole spinor fields were shown to be a legitimate solution of the Dirac field equation in a torsional setup \cite{esk,Fabbri:2010pk,Vignolo:2011qt}, whereas Elko and Majorana uncharged spinor fields represent type-5 spinors, although 
a recent example of a charged flagpole spinor has been shown to be a solution of the Dirac equation \cite{daRocha:2016bil}. More physical important examples on the Lounesto's classification can be 
found in Refs. \cite{daSilva:2012wp,alex}. It is worth to mention that  three  \textcolor{black}{further} classes of spinors that are beyond such a classification have been recently found, with ${\bf J}=0$, representing complementary flagpole spinors, pole spinors and flag spinors, with potential applications to provide genuinely quantum fields \cite{Villalobos:2015xca}. 

The characterization of exotic singular spinor fields in 
Lounesto's classes has introduced new fermions, including mass dimension one matter quantum fields, that have been studied in \cite{exotic}, with immediate physical applications to the problem of dark matter, after the works in  \cite{lee2,Fabbri:2010ws,Fabbri:2011mi}, with applications to the study of Hawking radiation \cite{bht}.  Other aspects of spinors fields classifications can be also found in Ref. \cite{mpla}.

The complex multivector field
${\rm Z}\in\sec\mathbb{C}\ell(M, g)$, where $\mathbb{C}\ell(M, g)$ denotes the
complexified spacetime Clifford bundle,  \textcolor{black}{is}
\begin{equation}
{\rm Z}(x)=\sigma(x)+\mathbf{J}(x)+i\mathbf{S}(x)+i(\mathbf{K}(x)+\omega(x))\gamma_{0123},
\label{boom1}
\end{equation}
where the bilinear covariants carry, respectively, the following unitary  irreducible representation of the Lorentz group:
\begin{table}[h]
\centering
\begin{tabular}[c]{||c|c||}
\hline\hline
Bilinear covariants &  Irreps. of the Lorentz group\\
\hline\hline
$\sigma(x)$&$ {(0,0)}$\\\hline
$\mathbf{J}(x)$&${\left(\f{1}{2},\f{1}{2}\right)}$\\\hline$\mathbf{S}(x)$& ${(1,0)\oplus(0,1)}$\\\hline
$\mathbf{K}(x)$&${\left(\f{1}{2},\f{1}{2}\right)}$\\\hline$\omega(x)$&${(0,0)}$\\\hline\hline
\end{tabular}   
%\caption{Bilinear covariants as the  irreducible representations of the Lorentz group.   }
\end{table}
\newpage

When the multivector operators $\sigma,\omega,\mathbf{J},\mathbf{S}%
,\mathbf{K}$ satisfy the Fierz identities, then the complex multivector
operator ${\rm Z}$ is denominated a {Fierz aggregate}.
Since the bilinear covariants are real, \textcolor{black}{they form} a multivector representation of $\mathbb{R}^{16}$. Hence, the Fierz identities for regular spinors describe a 7-dimensional submanifold wherein ${\rm Z}(x)$ resides, extending the Bloch sphere in Pauli formalism \cite{qtmosna}. When 
\beq\label{boome}
\gamma
_{0}{\rm Z}^{\dagger}(x)\gamma_{0} \,(=\bar{\rm Z}(x))={\rm Z}(x),\eeq which means that ${\rm Z}$ is a Dirac self-adjoint
aggregate, ${\rm Z}$ is called a {boomerang}.
A spinor field with either $\omega$ or $\sigma$  not equal to zero is said to be regular, whereas \textcolor{black}{when} $\omega=0=\sigma$  the spinor field \textcolor{black}{is said to be singular}. In this last case, the Fierz identities are superseded by \begin{subequations}\beq
\label{nilp}{\rm Z}^{2}(x)  &=&4\sigma(x) {\rm Z}(x),\\
{\rm Z}(x)\gamma_{\mu}{\rm Z}(x)&=&4J_{\mu}(x){\rm Z}(x),\\
{\rm Z}(x)i\gamma_{\mu\nu}{\rm Z}(x)&=&4S_{\mu\nu}(x){\rm Z}(x),\\
{\rm Z}(x)i\gamma_{0123}\gamma_{\mu}{\rm Z}(x)  &=&4K_{\mu}(x){\rm Z}(x),\\
 {\rm Z}(x)\gamma_{0123}{\rm Z}(x)&=&-4\omega(x) {\rm Z}(x).\label{nilp1}
\eeq\end{subequations}
\textcolor{black}{Henceforth in this section} the argument ``$x$'' in spinors and bilinears  is \textcolor{black}{left} implicit.
Any spinor field can be reconstructed from its
bilinear covariants, taking an arbitrary spinor field $\xi$
satisfying $\xi^{\dagger}\gamma_{0}\psi\neq0$: \begin{equation}
\psi=\frac{1}{4N}e^{-i\alpha}{\rm Z}\xi:=Z\xi, \label{a1}%
\end{equation}
\noindent where $N=\frac{1}{2}\sqrt{\xi^{\dagger}\gamma_{0}{\rm Z}\xi}$ and
$e^{-i\alpha}=\frac{1}{N}\xi^{\dagger}\gamma_{0}\psi$ \cite{Cra,Taka,qtmosna}.

\section{The problem of classifying second quantized spinors}

Our aim in this paper is to tackle the problem of classifying second quantized spinor fields. As pointed out in the introduction, the first question that arises is how to formulate the problem itself. In the classical (or perhaps, better, the first quantized) case, once we know the explicit expression of a spinor field we can compute
everything (energy, momentum, spin, etc.). In a second quantized theory things are not so straightforward: let us recall that solving a
second quantized theory amounts to computing all the correlators of the theory. A quantum reconstruction theorem exists, at least for ordinary
theories,  whose claim
is that (under certain general field theory axioms such as locality, hermiticity and Poincar\`e covariance) if we know all
the correlators of a theory we can reconstruct the fields themselves as well as their interactions \cite{wightman}. On the other
hand we have to consider also the other corner of the problem, i.e. the bilinear covariants of spinors. Bilinears, classically, are composites
of two spinor fields evaluated at the same spacetime point. Therefore the quantum correspondents must be regularized in
some way, for instance by normal ordering their bilinear expressions. Once this is accomplished, in analogy with the classical case, we
may ask if, knowing all the correlators of the bilinears, allows us to reconstruct the correlators of the spinor fields (perhaps up to a phase), because, in that case, the quantum reconstruction theorem would allow us to reconstruct the field themselves and their interactions. In this case the quantum classification of spinors could be reduced to the classification of the normal ordered bilinears. Unfortunately, although one such theorem is perhaps in the realm of possibilities, it is not in the realm of our capacities. Thus, for the time being, we have to conclude that we are not able to tackle the problem in such a generality. For the time being we content ourselves with a more modest, but more realistic program: a perturbative approach. 

An essential ingredient of perturbation theory in QFT is the free Feynman propagator (for in and out fields). It is based on the field expansion in terms of a complete set of (plane wave) solutions of the free field equation of motion.  Each term of the expansion has the form of a coefficient, which is either a creation or annihilation operator, times a plane wave, times a classical momentum dependent spinor. 
Plane waves satisfy the Klein-Gordon equation, so also the spinor field satisfies the same equation. In view of this therefore we remark that our spinors are very general, they are assumed to satisfy the Klein-Gordon equation (not necessarily the Dirac equation or other linear equations).
Let us focus now on the coefficient spinors. It is these classical spinors that play a central role in the
subsequent analysis. Being classical, we can apply to them the Lounesto classification scheme. However these spinors are momentum dependent, not coordinate dependent. Therefore the classification we apply to them is the classification in momentum space.  To be more explicit let us consider a classical spinor $\psi(x)$, in a flat Minkowski spacetime, which admits a Fourier transform 
\beq
\psi(p) = \int d^4x\,\psi(x) \, e^{-ipx}.
\eeq 
In the same way as we form the bilinear covariants $\bar \psi(x) \Gamma_i \psi(x)$ in coordinate space, we can form also the bilinears in momentum space, $\bar \psi(p) \Gamma_i \psi(p)$. The latter satisfy formally the same Fierz identities as in coordinate space, since the Fierz identities are purely algebraic relations. Therefore in momentum space the same classification holds as in coordinate space (Lounesto). However the two classifications are unrelated, because the bilinears in momentum space are not the Fourier transform of the bilinears in coordinate space, and the Fierz identities in momentum space are not the Fourier transform of the Fierz identities in coordinate space, and vice versa. The relation between the two classifications is a problem we leave for the future. The important point we would like to stress is that, assuming the usual plane wave expansion where the spinors, used as expansion coefficients, can run through all the regular and singular spinors in the classical classification, we show that the second quantized regular and singular spinors are split into additional subclasses of fields.

In the next section we carry out the analysis of the scalar and pseudoscalar bilinears constructed out of spinor fields located (in general) at different spacetime points.  Of course the analysis should be extended also to the other bilinear covariants. But these two are in a sense basic and, anyhow, enough to appreciate the novelties of the second quantized classification.

\section{Computing the second quantized scalar and pseudoscalar bilinear covariants}
 
From a practical point of view, the computation of the scalar and pseudoscalar bilinear covariants are necessary in order to perform a sufficient investigation on the nature of the spinor at hand. In fact, their values select regular (when at least one of them is different from zero) and singular (otherwise) spinors.  In this section we shall focus on them.

Recall that the standard Dirac quantum field, constructed upon standard Dirac spinors, is well known to read
 \beq\label{psi}
 \psi(x)=\int \frac{d^3{\bf p}}{(2\pi)^3\sqrt{2E_{\bf p}}}\sum_{s=1,2}\;\left(a_{{\bf p},s} u^s(p)\,e^{-ip\cdot{x}}+b^\dagger_{{\bf p},s} v^s(p)\,e^{i{p}\cdot{x}}\right),\eeq where, given the Pauli matrices $\sigma^i$ and $\sigma = (\sigma^1,\sigma^2,\sigma^3)$,   
    \beq
    u^s(p)&=&\binom{\sqrt{p\cdot \sigma}\;\zeta^s}{\sqrt{p\cdot \bar\sigma}\;\zeta^s},\qquad
    v^s(p)=\binom{\sqrt{p\cdot \sigma}\;\eta^s}{\sqrt{-p\cdot \bar\sigma}\;\eta^s},\quad s=1(=+),2(=-),
    \eeq
   and 
    $
    \zeta^1=\binom{1}{0} =\eta^1, \zeta^2=\eta^2=\binom{0}{1}$ constitute an orthogonal 2-spinor basis. Dirac spinors are solutions of the massive Dirac equation and a superposition of plane waves, satisfying 
\beq\label{uv}
(\slash{\!\!\!p}-m)u_r(p)=0,\qquad\qquad
(\slash{\!\!\!p}+m)v_s(p)=0,
\eeq\noindent and $\bar{u}_r(p)v_s(p)=0$. Moreover, the dual field is constructed as 
\beq\label{barpsi}
 \bar\psi(x)=\int \frac{d^3{\bf p}}{(2\pi)^3\sqrt{2E_{\bf p}}}\sum_{s=1,2}\;\left(a^\dagger_{{\bf p},s} \bar{u}^s(p)\,e^{ip\cdot{x}}+b_{{\bf p},s} \bar{v}^s(p)\,e^{-i{p}\cdot{x}}\right).\eeq

To evaluate the second quantized bilinear covariants, quantum fields are assumed to be constructed upon 
general expansion coefficients, 
   \beq\label{psigen}
 \uppsi(x)&=&\int \frac{d^3{\bf p}}{(2\pi)^3\sqrt{2E_{\bf p}}}\sum_{s=1,2}\;\left(a_{{\bf p},s} \uppsi_1^s(p)\,e^{-ip\cdot{x}}+b^\dagger_{{\bf p},s} \uppsi_2^s(p)\,e^{i{p}\cdot{x}}\right),\\
 \bar\uppsi(x)&=&\int \frac{d^3{\bf p}}{(2\pi)^3\sqrt{2E_{\bf p}}}\sum_{s=1,2}\;\left(b_{{\bf p},s} \bar{\uppsi}_2^s(p)\,e^{-i{p}\cdot{x}}+a^\dagger_{{\bf p},s} \bar{\uppsi}_1^s(p)\,e^{ip\cdot{x}}\right),\label{psigendual}\eeq
where $\pwpw$ and $\pwpe$ are arbitrary spinors in the Lounesto's classification. 
 
Although the spinors in (\ref{psi}) and (\ref{barpsi}) could be imposed to be in the same Lounesto's spinor class (\ref{tipo1} -- \ref{tipo6}), there is no reason, a priori, to preclude the possibility that the spinors $\uppsi_1$ and $\uppsi_2$ are not in the same spinor class. In fact, the Dirac equation has solutions beyond the well known textbook eigenspinors of the parity operator in the class 1 of the Lounesto's classification, further encoding also flag-dipole type-4  \cite{esk} and flagpole type-5 \cite{daRocha:2016bil} spinor solutions of the Dirac equation. Besides, one considers, as usual, anti-commutators among creation/annihilation operators
\beq
\{a_{{\bf p},s}, a^\dagger_{{\bf p}',s'}\}=(2\pi)^3\delta^{(3)}({\bf p}-{\bf p}')\delta^{ss'}=\{b_{{\bf p},s}, b^\dagger_{{\bf p}',s'}\}\,,\eeq\noindent with all other anti-commutators  equal to  zero.
    
By the reconstruction theorem \cite{Taka,lou2}, one can consider Eq. (\ref{a1}) to yield
    \beq\label{recc}
    \pwpw^s(p)=Z_1(p)\,\xi^s_1(p),\qquad\qquad
     \pwpe^s(p)=Z_2(p)\,\xi^s_2(p),\eeq
 denoting $Z_s=Z_s(p)$ hereon.
    The next subsection is devoted to use the general quantum fields (\ref{psigen}, \ref{psigendual}) to construct the bilinear covariants in the second quantization. 
    
%----------------------------------------------    

    \subsection{Scalar bilinear covariant}
First we shall compute the scalar bilinear covariant
    \beq\label{2ndsigma123}
   \upsigma(x)&=&\,\wpwp(x)\;\pwp(x) .
   \eeq
   However, for further use of terms of type $\wpwp(x)\;\pwp(x')$ in Lagrangians and in  the calculation of propagators, we want to compute the most general covariant quantity $\wpwp(x)\;\pwp(x')$. A posteriori one can make $x=x'$ and reobtain Eq. (\ref{2ndsigma123}). Hence, 
\ba\label{2ndsigma}
 \wpwp(x)\pwp(x')&=&\left. \int \frac{d^3{\bf p}d^3{\bf p}'}{(2\pi)^6 2\sqrt{E_{\bf p}E_{{\bf p}'}}} \sum_{s,s'=1,2}\Bigg[\overbrace{b_{{\bf p},s}a_{{\bf p}',s'}\bar\pwp_2^s(p)\pwpw^{s'}\!(p')e^{-i(p\cdot x+p'\cdot x')}}^{\mathbb{1}}+\overbrace{b_{{\bf p},s}b^\dagger_{{\bf p}',s'}\bar\pwp_2^s(p)\pwpe^{s'}\!(p')e^{i(p'\cdot x'-p\cdot x)}}^{\mathbb{2}}\right.\nonumber\\&+&\left.\overbrace{a^\dagger_{{\bf p},s}a_{{\bf p'},s'}\bar\pwp_1^s(p)\pwp_1^{s'}\!(p')}^{\mathbb{3}}e^{i(p\cdot x - p'\cdot x')}+\overbrace{a^\dagger_{{\bf p},s}b^\dagger_{{\bf p}',s'}\bar\pwp_1^s(p)\uppsi_2^{s'}\!(p')e^{i(p\cdot x+p'\cdot x')}}^{\mathbb{4}} \Bigg]\right. . 
\ea

Let us, then, analyze the second quantized scalar bilinear covariant $\upsigma$. Eq. (\ref{2ndsigma}) has four terms that shall be scrutinized. When $\uppsi_1$ and $\uppsi_2$ are a set 
   of regular spinors, $\upsigma\neq0$, since the spinor content of all terms $\mathbb{1} - {\mathbb{4}}$ in Eq. (\ref{2ndsigma}) is not equal to zero.    
   When all such terms equal zero, then obviously $ \upsigma=0$. However, at most one can suppose that $\{\uppsi_1, \uppsi_2\}$ is a set of singular spinors. In this case, the terms ${\mathbb{2}}$ and ${\mathbb{3}}$ in Eq. (\ref{2ndsigma}) can be equal to zero just when $p=p'$, otherwise such terms are not equal to zero. Even when $p=p'$, the mixed terms  ${\mathbb{1}}$ and ${\mathbb{4}}$ do not necessarily vanish. 
   Hence, a set $\{\uppsi_1, \uppsi_2\}$ of singular spinors can generate  a quantum field that is not singular. To construct singular quantum fields from singular spinors, the conditions  
$\bar\pwp_2^s(p)\pwpw^{s'}(p')=0= \bar\pwp_1^s(p)\uppsi_2^{s'}(p')$ must be imposed, at $p=p'$.  
   A more refined analysis arises when the reconstruction theorem (\ref{a1}, \ref{recc}) is taken into account and this shall be our main focus in what follows, with explicit and disjoint possibilities:
   \begin{enumerate}
    \item[1)] In the following calculations we consider just the core spinor part of the $\mathbb{1}$ term, $b_{{\bf p},s}a_{{\bf p}',s'}\bar\pwp_2^s(p)\pwpw^{s'}(p')$, in Eq. (\ref{2ndsigma}), namely, \beq\bar\pwp_2^s(p)\pwpw^{s'}(p').  \label{ppli}  \eeq 
    The reconstruction theorem yields,
     \beq\bar\pwp_2^s(p)\pwpw^{s'}(p')=    \bar\xi_2^s\bar{Z_2}(p)Z_1(p')\xi_1^{s'}.\label{z1z2}\eeq

    Since $\bar{Z_2}Z_1$ is, in general, a multivector, then to analyze whether the term $\bar\pwp_2^s(p)\pwpw^{s'}(p')$ is null resides on the scrutiny of the product $\bar{Z_2}Z_1$. In what follows, the bilinear covariants indexed by ${}_{\scriptsize 1}$ refers to 
the argument $(p')$, whereas the ones indexed by  ${}_{\scriptsize 2}$ refers to 
the argument $(p)$. 
    The definition of the boomerang in Eqs. (\ref{boom1}, \ref{boome}) yields $
    \bar{Z}_a=Z_a$. In fact, the homogeneous parts of (\ref{boome}) correspond to the bilinear covariants of some spinor \cite{lou2}. 
    Let us calculate, thus, Eq. (\ref{z1z2})
    from the aggregates for the spinors $\pwpw$ and $\pwpe$, respectively:
\beq
{\rm Z}_1&=&\sigma_1+\mathbf{J}_1+i\mathbf{S}_1+i\mathbf{K}_1\gamma_{0123}+\omega_1\gamma_{0123},\qquad
{\rm Z}_2=\sigma_2+\mathbf{J}_2+i\mathbf{S}_2+i\mathbf{K}_2\gamma_{0123}+\omega_2\gamma_{0123}.    \label{z2}
\eeq
Hence, employing these two equations for the boomerangs for the spinors $\pwpw$ and $\pwpe$  yields the following complex multivector:
    \beq\label{z22}
    {\rm Z}_2{\rm Z}_1 &=&(\sigma_1+\mathbf{J}_1+i\mathbf{S}_1+i\mathbf{K}_1\gamma_{0123}+\omega_1\gamma_{0123})(\sigma_2+\mathbf{J}_2+i\mathbf{S}_2+i\mathbf{K}_2\gamma_{0123}+\omega_2\gamma_{0123})\nonumber\\&=&\sigma_1\sigma_2+\mathbf{J}_1\mathbf{J}_2-\mathbf{S}_1\mathbf{S}_2+\mathbf{K}_1\mathbf{K}_2+\omega_1\omega_2
    +\sigma_1\mathbf{J}_2+\sigma_2\mathbf{J}_1+(\sigma_1\omega_2+\omega_1\sigma_2)\gamma_{0123}\nonumber\\&&
    +(\mathbf{J}_1\omega_2+\mathbf{J}_2\omega_1)\gamma_{0123}
-(\mathbf{S}_1\mathbf{K}_2+\mathbf{S}_2\mathbf{K}_1)\gamma_{0123}+(\mathbf{S}_1\omega_2+\mathbf{S}_2\omega_1)\gamma_{0123}+(\mathbf{K}_1\omega_2+\mathbf{K}_2\omega_1)\gamma_{0123}
\nonumber\\
&&+i\left[\sigma_1\mathbf{S}_2+\sigma_2\mathbf{S}_1+(\sigma_1\mathbf{K}_2+\sigma_2\mathbf{K}_1)\gamma_{0123}+\mathbf{J}_1\mathbf{S}_2+\mathbf{J}_2\mathbf{S}_1  -(\mathbf{J}_1\mathbf{K}_2+\mathbf{J}_2\mathbf{K}_1)\gamma_{0123}\right].\eeq The above expression, Eq. (\ref{z22}), shall be now analyzed to verify
all the possibilities to make the term (\ref{ppli}) to be null. 
First, one should split the Clifford products, 
\begin{subequations}\beq
\mathbf{J}_1\mathbf{K}_2&=&
\mathbf{J}_1\lrcorner\mathbf{K}_2+
\mathbf{J}_1\wedge\mathbf{K}_2, \quad\qquad\qquad
\mathbf{S}_1\mathbf{K}_2= \mathbf{S}_1\lrcorner\mathbf{K}_2+
\mathbf{S}_2\wedge\mathbf{K}_1,\\
\mathbf{J}_1\mathbf{J}_2&=&\mathbf{J}_1\lrcorner\mathbf{J}_2
+\mathbf{J}_1\wedge\mathbf{J}_2, \quad\quad
\mathbf{S}_1\mathbf{S}_2=\mathbf{S}_1\lrcorner\mathbf{S}_2+\mathbf{S}_1\wedge\mathbf{S}_2,\quad\quad 
\mathbf{K}_1\mathbf{K}_2=\mathbf{K}_1\lrcorner\mathbf{K}_2+\mathbf{K}_1\wedge\mathbf{K}_2,
\eeq that subsequently hold for the interchange $1\leftrightarrow2$. 
\end{subequations}
\item[1.1)] When the spinors $\pwpw, \pwpe$ are regular spinors, according to Lounesto standard classification, it implies that all terms in Eq. (\ref{2ndsigma}) do not equal zero.
\item[1.2)] When the spinors $\pwpw, \pwpe$ are both singular spinors, it means that $\sigma_1=\omega_1=0=\sigma_2=\omega_2$. Hence, Eq. (\ref{z22}) reads
 \beq\label{z00}
    {\rm Z}_2{\rm Z}_1 &=&\mathbf{J}_1\mathbf{J}_2-\mathbf{S}_1\mathbf{S}_2+\mathbf{K}_1\mathbf{K}_2-(\mathbf{S}_1\mathbf{K}_2+\mathbf{S}_2\mathbf{K}_1)\gamma_{0123}
+i\left[\mathbf{J}_1(\mathbf{S}_2-\mathbf{K}_2\gamma_{0123}
)+\mathbf{J}_2(\mathbf{S}_1-\mathbf{K}_1\gamma_{0123}
)\right].\eeq
In order to have $ {\rm Z}_2{\rm Z}_1=0$ in this case, 
both the real and the complex part must equal zero. Hence, we shall 
scrutinize the subcases that follows from this case 1.2):
\begin{enumerate}
\item[1.2.1)] For the case where $\pwpw,\pwpe$ are both type-4, flag-dipole spinors in the Lounesto's  classification, one has, according to Eq. (\ref{tipo4}), the values for the bilinear covariants $
\sigma_a=0=\omega_a, \mathbf{K}_a\neq0$ and $\mathbf{S}_a\neq0$, for $a=1,2$.  
Now, let us consider Eq. (\ref{z00}) and derive what are the conditions that make the complex multivector ${\rm Z}_2{\rm Z}_1$ to equal zero. For it, let us split ${\rm Z}_2{\rm Z}_1$ in Eq. (\ref{z00}) into its non-zero homogeneous parts, 
\begin{subequations}
\beq\label{ssqq}
&&\!\!\!\!\!\!\!\!\!\!\!\!\!\!\!\!\!\!\mathbf{J}_1\lrcorner\mathbf{J}_2-\mathbf{S}_1\lrcorner\mathbf{S}_2+\mathbf{K}_1\lrcorner\mathbf{K}_2
\in\sec\Omega_{}^0(M),\\
&&\!\!\!\!\!\!\!\!\!\!\!\!\!\!\!\!\!\!(\mathbf{S}_1\wedge\mathbf{K}_2+\mathbf{S}_2\wedge\mathbf{K}_1)\gamma_{0123}\in\sec\Omega_{}^1(M),\\
&&\!\!\!\!\!\!\!\!\!\!\!\!\!\!\!\!\!\mathbf{J}_1\!\wedge\!\mathbf{J}_2\!+\!\mathbf{K}_1\!\wedge\!\mathbf{K}_2\!+\!\langle\mathbf{S}_1\mathbf{S}_2\rangle_2
\!-\!i(\mathbf{J}_1\!\wedge\!\mathbf{K}_2\!+\!\mathbf{J}_2\!\wedge\!\mathbf{K}_1)\gamma_{0123}\in\sec\Omega_{}^2(M),\\
&&\!\!\!\!\!\!\!\!\!\!\!\!\!\!\!\!\!\!(\mathbf{S}_1\llcorner\mathbf{K}_2-\mathbf{S}_2\llcorner\mathbf{K}_1)\gamma_{0123}+i(\mathbf{J}_1\wedge\mathbf{S}_2-\mathbf{J}_2\wedge\mathbf{S}_1)\;\;\in\sec\Omega_{}^3(M),\\
&&\!\!\!\!\!\!\!\!\!\!\!\!\!\!\!\!\!\!-\mathbf{S}_1\wedge\mathbf{S}_2-i\left[\mathbf{J}_1\lrcorner\mathbf{K}_2+\mathbf{J}_2\lrcorner\mathbf{K}_1\right]\gamma_{0123}\;\in\sec\Omega_{}^4(M),
\eeq
\end{subequations}
and equal them to zero. 
In order to verify which are the conditions that the bilinear covariants must satisfy to force ${\rm Z}_2{\rm Z}_1$ to be zero, one must equal Eqs. (\ref{ssqq}) to zero, yielding the following simultaneous conditions:
\begin{subequations}
\beq\label{cflagdi}
\mathbf{J}_1\lrcorner\mathbf{J}_2-\mathbf{S}_1\lrcorner\mathbf{S}_2+\mathbf{K}_1\lrcorner\mathbf{K}_2
&=&0=
\mathbf{S}_1\wedge\mathbf{K}_2+\mathbf{S}_2\wedge\mathbf{K}_1,\\
\mathbf{J}_1\!\wedge\!\mathbf{J}_2\!+\!\mathbf{K}_1\!\wedge\!\mathbf{K}_2\!+\!\langle\mathbf{S}_1\mathbf{S}_2\rangle_2&=&0=\mathbf{J}_1\!\wedge\!\mathbf{K}_2\!+\!\mathbf{J}_2\!\wedge\!\mathbf{K}_1=
\mathbf{S}_1\wedge\mathbf{S}_2=\mathbf{J}_1\lrcorner\mathbf{K}_2+\mathbf{J}_2\lrcorner\mathbf{K}_1\,.
\label{418}
\eeq
\end{subequations}
When Eqs. (\ref{cflagdi}) - (\ref{418}) hold, it yields the second quantized scalar bilinear covariant $\upsigma = \wpwp\;\pwp$ to be null.
\item[1.2.2)] For the case where $\pwpw,\pwpe$ are both type-5, flagpole spinors in the Lounesto's classification, one has, according to Eq. (\ref{tipo5}), the values for the bilinear covariants $
\sigma_a=0=\omega_a, \;\mathbf{K}_a=0,\; \mathbf{S}_a\neq0.$ 
Now, let us consider Eq. (\ref{z00}) and see what are the conditions that make the complex multivector ${\rm Z}_2{\rm Z}_1$ to equal zero. For it, the expression for ${\rm Z}_2{\rm Z}_1$ in Eq. (\ref{z00}) must split into its non-zero homogeneous parts,
\begin{subequations}
\beq\label{ssq1}
&&\!\!\!\!\!\!\!\!\!\!\!\!\!\!\!\!\!\!\mathbf{J}_1\lrcorner\mathbf{J}_2-\mathbf{S}_1\lrcorner\mathbf{S}_2\in\sec\Omega_{}^0(M),\\
&&\!\!\!\!\!\!\!\!\!\!\!\!\!\!\!\!\!\!\mathbf{J}_1\!\wedge\!\mathbf{J}_2\!+\!\langle\mathbf{S}_1\mathbf{S}_2\rangle_2
\in\sec\Omega_{}^2(M),\\
&&\!\!\!\!\!\!\!\!\!\!\!\!\!\!\!\!\!\!i(\mathbf{J}_1\wedge\mathbf{S}_2-\mathbf{J}_2\wedge\mathbf{S}_1)\;\;\in\sec\Omega_{}^3(M),\\
&&\!\!\!\!\!\!\!\!\!\!\!\!\!\!\!\!\!\!-\mathbf{S}_1\wedge\mathbf{S}_2\gamma_{0123}\;\in\sec\Omega_{}^4(M).\label{ssq11}
\eeq
\end{subequations}
Now, to verify which are the conditions that the bilinear covariants must satisfy to force ${\rm Z}_2{\rm Z}_1$ to be zero, we must equal Eqs. (\ref{ssq1}) - (\ref{ssq11}) to zero. It yields the following simultaneous conditions:
\beq\label{cflag}
\mathbf{J}_1\lrcorner\mathbf{J}_2-\mathbf{S}_1\lrcorner\mathbf{S}_2
=0= 
\mathbf{J}_1\!\wedge\!\mathbf{J}_2\!+\!\langle\mathbf{S}_1\mathbf{S}_2\rangle_2=
\mathbf{S}_1\wedge\mathbf{S}_2.
\label{419}
\eeq
When Eqs. (\ref{cflag}) hold, then that the second quantized scalar bilinear covariant is null.
\item[1.2.3)] For the case where $\pwpw,\pwpe$ are both type-6, dipole spinors in the Lounesto's classification, one has, according to Eq. (\ref{tipo4}), $
\sigma_a=0=\omega_a, \;\mathbf{K}_a\neq0,\;\mathbf{S}=0.$ Consider Eq. (\ref{z00}), the conditions that make the complex multivector ${\rm Z}_2{\rm Z}_1$ to be equal to  zero shall be derived, by splitting ${\rm Z}_2{\rm Z}_1$ in Eq. (\ref{z00}) into its non-zero homogeneous parts,
\begin{subequations}
\beq\label{ssq6}
&&\!\!\!\!\!\!\!\!\!\!\!\!\!\!\!\!\!\!\mathbf{J}_1\lrcorner\mathbf{J}_2+\mathbf{K}_1\lrcorner\mathbf{K}_2
\in\sec\Omega_{}^0(M),\\
&&\!\!\!\!\!\!\!\!\!\!\!\!\!\!\!\!\!\!\mathbf{J}_1\!\wedge\!\mathbf{J}_2\!+\!\mathbf{K}_1\!\wedge\!\mathbf{K}_2\!-\!i(\mathbf{J}_1\!\wedge\!\mathbf{K}_2\!+\!\mathbf{J}_2\!\wedge\!\mathbf{K}_1)\gamma_{0123}\in\sec\Omega_{}^2(M),\\
&&\!\!\!\!\!\!\!\!\!\!\!\!\!\!\!\!\!\!-i\left[\mathbf{J}_1\lrcorner\mathbf{K}_2+\mathbf{J}_2\lrcorner\mathbf{K}_1\right]\gamma_{0123}\;\in\sec\Omega_{}^4(M).\label{ssq61}
\eeq
\end{subequations} Following the same reasoning of the previous analyses, the conditions to be satisfied to have ${\rm Z}_2{\rm Z}_1=0$ yields to the following set of simultaneous conditions: 
\beq\label{cflagdip}
\mathbf{J}_1\lrcorner\mathbf{J}_2+\mathbf{K}_1\lrcorner\mathbf{K}_2=0=
\mathbf{J}_1\!\wedge\!\mathbf{J}_2\!+\!\mathbf{K}_1\!\wedge\!\mathbf{K}_2=\mathbf{J}_1\!\wedge\!\mathbf{K}_2\!+\!\mathbf{J}_2\!\wedge\!\mathbf{K}_1
\mathbf{J}_1\lrcorner\mathbf{K}_2+\mathbf{J}_2\lrcorner\mathbf{K}_1.
\label{420}
\eeq
Eqs. (\ref{cflagdip}) then imply the second quantized scalar bilinear covariant to be null.
\item[1.2.4)] For the case where $\pwp_2$ is a regular spinor, of type-1, and $\pwp_1$ is a type-5 spinor (obviously these roles are interchangeable, $\pwpw\leftrightarrow\pwpe$), it follows that  \beq\label{z24}
   \!\!\!\!\! {\rm Z}_2{\rm Z}_1\! =\!\mathbf{J}_1\mathbf{J}_2\!-\!\mathbf{S}_1\mathbf{S}_2\!+\!\sigma_2\mathbf{J}_1
    \!+\!\mathbf{J}_1\omega_2\gamma_{0123}
\!-\!\mathbf{S}_1\mathbf{K}_2\gamma_{0123}
\!+\!i\left[\sigma_2\mathbf{S}_1\!+\!\mathbf{J}_1\mathbf{S}_2\!+\!\mathbf{J}_2\mathbf{S}_1  -\mathbf{J}_1\mathbf{K}_2\gamma_{0123}+\mathbf{S}_1\omega_2\gamma_{0123}\right].\eeq
By splitting this complex multivector into its homogeneous 
parts we have
\begin{subequations}
\beq\label{ssqq11}
&&\!\!\!\!\!\!\!\!\!\!\!\!\!\!\!\!\!\!\mathbf{J}_1\lrcorner\mathbf{J}_2-\mathbf{S}_1\lrcorner\mathbf{S}_2
\in\sec\Omega_{}^0(M),\\
&&\!\!\!\!\!\!\!\!\!\!\!\!\!\!\!\!\!\!\sigma_2\mathbf{J}_1+\mathbf{S}_1\wedge\mathbf{K}_2\gamma_{0123}\in\sec\Omega_{}^1(M),\\
&&\!\!\!\!\!\!\!\!\!\!\!\!\!\!\!\!\!\!\mathbf{J}_1\!\wedge\!\mathbf{J}_2\!+\!\langle\mathbf{S}_1\mathbf{S}_2\rangle_2\!+\mathbf{S}_1\omega_2\gamma_{0123}\!
\!-\!i(\sigma_2\mathbf{S}_1+\mathbf{J}_1\!\wedge\!\mathbf{K}_2\gamma_{0123})\in\sec\Omega_{}^2(M),\\
&&\!\!\!\!\!\!\!\!\!\!\!\!\!\!\!\!\!\!(\mathbf{J}_1\omega_2+\mathbf{S}_1\llcorner\mathbf{K}_2)\gamma_{0123}+i(\mathbf{J}_1\wedge\mathbf{S}_2-\mathbf{J}_2\wedge\mathbf{S}_1)\;\;\in\sec\Omega_{}^3(M),\\
&&\!\!\!\!\!\!\!\!\!\!\!\!\!\!\!\!\!\!-\mathbf{S}_1\wedge\mathbf{S}_2-i\mathbf{J}_1\lrcorner\mathbf{K}_2\gamma_{0123}\;\in\sec\Omega_{}^4(M).
\eeq
\end{subequations}
Hence, for  the condition ${\rm Z}_2{\rm Z}_1=0$ to hold, the following equations must be simultaneously satisfied:
\beq\label{ssqq13}
\mathbf{J}_1\lrcorner\mathbf{J}_2-\mathbf{S}_1\lrcorner\mathbf{S}_2=0=
\sigma_2\mathbf{J}_1+\mathbf{S}_1\wedge\mathbf{K}_2\gamma_{0123}=
\mathbf{J}_1\!\wedge\!\mathbf{J}_2\!+\!\langle\mathbf{S}_1\mathbf{S}_2\rangle_2\!+\mathbf{S}_1\omega_2\gamma_{0123}=\sigma_2\mathbf{S}_1+\mathbf{J}_1\!\wedge\!\mathbf{K}_2\gamma_{0123},\\
(\mathbf{J}_1\omega_2+\mathbf{S}_1\llcorner\mathbf{K}_2)\gamma_{0123}=0=\mathbf{J}_1\wedge\mathbf{S}_2-\mathbf{J}_2\wedge\mathbf{S}_1=
\mathbf{S}_1\wedge\mathbf{S}_2=\mathbf{J}_1\lrcorner\mathbf{K}_2\gamma_{0123}.
\eeq
\item[1.2.5)] When $\pwpe$ is a regular spinor and $\pwpw$ is a type-6, using Eq. (\ref{tipo6}) yields 
    \beq\label{z26}
      \!\!\!\!\!\!\!\! \!\!\!\!\!\!\!\!\!\!\!   {\rm Z}_2{\rm Z}_1\!=\!\mathbf{J}_1\mathbf{J}_2\!+\!\mathbf{K}_1\mathbf{K}_2
 \!+\!\sigma_2\mathbf{J}_1
    \!+\!\mathbf{J}_1\omega_2\gamma_{0123}
\!-\!\mathbf{S}_2\mathbf{K}_1\gamma_{0123}
\!+\!i\left[\sigma_2\mathbf{K}_1\gamma_{0123}\!+\!\mathbf{J}_1\mathbf{S}_2 \!-\!(\mathbf{J}_1\mathbf{K}_2\!+\!\mathbf{J}_2\mathbf{K}_1\!+\!\mathbf{K}_1\omega_2)\gamma_{0123}\right],\eeq whose splitting into its homogeneous 
parts reads 
\begin{subequations}
\beq\label{ssq66}
&&\!\!\!\!\!\!\!\!\!\!\!\!\!\!\!\!\!\!\!\!\!\!\mathbf{J}_1\lrcorner\mathbf{J}_2+\mathbf{K}_1\lrcorner\mathbf{K}_2
\in\sec\Omega_{}^0(M),\\
&&\!\!\!\!\!\!\!\!\!\!\!\!\!\!\!\!\!\!\!\!\!\!\sigma_2\mathbf{J}_1+\mathbf{S}_2\wedge\mathbf{K}_1\gamma_{0123}\in\sec\Omega_{}^1(M),\\
&&\!\!\!\!\!\!\!\!\!\!\!\!\!\!\!\!\!\!\!\!\!\!\mathbf{J}_1\!\wedge\!\mathbf{J}_2\!+\!\mathbf{K}_1\!\wedge\!\mathbf{K}_2\!-\!i(\mathbf{J}_1\!\wedge\!\mathbf{K}_2\!+\!\mathbf{J}_2\!\wedge\!\mathbf{K}_1)\gamma_{0123}\in\sec\Omega_{}^2(M),\\
&&\!\!\!\!\!\!\!\!\!\!\!\!\!\!\!\!\!\!\!\!\!\!\star(-\!\mathbf{S}_2\llcorner\mathbf{K}_1\!+\!(\mathbf{K}_1\!+\!\mathbf{J}_1)\omega_2)\!+\!i(\sigma_2\mathbf{K}_1\gamma_{0123}\!+\!\mathbf{J}_1\wedge\mathbf{S}_2)\in\sec\Omega_{}^3(M),\\
&&\!\!\!\!\!\!\!\!\!\!\!\!\!\!\!\!\!\!\!\!\!\!-\mathbf{S}_1\wedge\mathbf{S}_2-i\left[\mathbf{J}_1\lrcorner\mathbf{K}_2+\mathbf{J}_2\lrcorner\mathbf{K}_1\right]\gamma_{0123}\;\in\sec\Omega_{}^4(M).
\eeq
\end{subequations}
Hence, the condition ${\rm Z}_2{\rm Z}_1=0$ implies
\begin{subequations}
\beq\label{q66}
\!\!\!\!\!\!\mathbf{J}_1\lrcorner\mathbf{J}_2+\mathbf{K}_1\lrcorner\mathbf{K}_2=0=
\sigma_2\mathbf{J}_1+\mathbf{S}_2\wedge\mathbf{K}_1\gamma_{0123}=
\mathbf{J}_1\!\wedge\!\mathbf{J}_2\!+\!\mathbf{K}_1\!\wedge\!\mathbf{K}_2=\mathbf{J}_1\!\wedge\!\mathbf{K}_2\!+\!\mathbf{J}_2\!\wedge\!\mathbf{K}_1,\\
-\!\mathbf{S}_2\llcorner\mathbf{K}_1\!+\!(\mathbf{K}_1\!+\!\mathbf{J}_1)\omega_2=0=\sigma_2\mathbf{K}_1\gamma_{0123}\!+\!\mathbf{J}_1\!\wedge\!\mathbf{S}_2\!=
\mathbf{J}_1\lrcorner\mathbf{K}_2+\mathbf{J}_2\lrcorner\mathbf{K}_1.
\eeq
\end{subequations}

\item[1.2.6)] Finally, for the case where $\pwpw$ is a regular spinor and $\pwpe$ is a type-4 spinor, one has, according to Eq. (\ref{tipo4}) 
       \beq\label{z28}
   \!\!\!\!\!\!\!\! {\rm Z}_2{\rm Z}_1 &=&\mathbf{J}_1\mathbf{J}_2-\mathbf{S}_1\mathbf{S}_2+\mathbf{K}_1\mathbf{K}_2+\sigma_2\mathbf{J}_1
    +\mathbf{J}_1\omega_2\gamma_{0123}
-(\mathbf{S}_1\mathbf{K}_2+\mathbf{S}_2\mathbf{K}_1)\gamma_{0123}\nonumber
\\
&&+i\left[\sigma_2\mathbf{S}_1+\sigma_2\mathbf{K}_1\gamma_{0123}+\mathbf{J}_1\mathbf{S}_2+\mathbf{J}_2\mathbf{S}_1  -(\mathbf{J}_1\mathbf{K}_2+\mathbf{J}_2\mathbf{K}_1)\gamma_{0123}+\mathbf{S}_1\omega_2\gamma_{0123}+\mathbf{K}_1\omega_2\gamma_{0123}\right].\eeq
Splitting  into homogeneous 
parts, 
\begin{subequations}
\beq\label{ssq67}
&&\!\!\!\!\!\!\!\!\!\!\!\!\!\!\!\!\!\!\!\!\!\!\mathbf{J}_1\lrcorner\mathbf{J}_2-\mathbf{S}_1\lrcorner\mathbf{S}_2+\mathbf{K}_1\lrcorner\mathbf{K}_2
\in\sec\Omega_{}^0(M),\\
&&\!\!\!\!\!\!\!\!\!\!\!\!\!\!\!\!\!\!\!\!\!\!\sigma_2\mathbf{J}_1+(\mathbf{S}_1\wedge\mathbf{K}_2+\mathbf{S}_2\wedge\mathbf{K}_1)\gamma_{0123}\in\sec\Omega_{}^1(M),\\
&&\!\!\!\!\!\!\!\!\!\!\!\!\!\!\!\!\!\!\!\!\!\!\mathbf{J}_1\!\wedge\!\mathbf{J}_2\!+\!\mathbf{K}_1\!\wedge\!\mathbf{K}_2\!+\!\langle\mathbf{S}_1\mathbf{S}_2\rangle_2
\!-\!i(\mathbf{J}_1\!\wedge\!\mathbf{K}_2\!+\!\mathbf{J}_2\!\wedge\!\mathbf{K}_1)\gamma_{0123}\in\sec\Omega_{}^2(M),\\
&&\!\!\!\!\!\!\!\!\!\!\!\!\!\!\!\!\!\!\!\!\!\!\star(\mathbf{S}_1\llcorner\mathbf{K}_2\!-\!\mathbf{S}_2\llcorner\mathbf{K}_1\!+\!(\mathbf{K}_1\!+\!\mathbf{J}_1)\omega_2)\!+\!i(\sigma_2\mathbf{K}_1\gamma_{0123}\!+\!\mathbf{J}_1\wedge\mathbf{S}_2\!-\!\mathbf{J}_2\wedge\mathbf{S}_1)\in\sec\Omega_{}^3(M),\\
&&\!\!\!\!\!\!\!\!\!\!\!\!\!\!\!\!\!\!\!\!\!\!-\mathbf{S}_1\wedge\mathbf{S}_2-i\left[\mathbf{J}_1\lrcorner\mathbf{K}_2+\mathbf{J}_2\lrcorner\mathbf{K}_1\right]\gamma_{0123}\;\in\sec\Omega_{}^4(M),
\eeq
\end{subequations}
the condition ${\rm Z}_2{\rm Z}_1=0$ yields
\begin{subequations}
\beq\label{q66}
\mathbf{J}_1\lrcorner\mathbf{J}_2-\mathbf{S}_1\lrcorner\mathbf{S}_2+\mathbf{K}_1\lrcorner\mathbf{K}_2=0=
\sigma_2\mathbf{J}_1+(\mathbf{S}_1\wedge\mathbf{K}_2+\mathbf{S}_2\wedge\mathbf{K}_1)\gamma_{0123}=
\mathbf{J}_1\!\wedge\!\mathbf{J}_2\!+\!\mathbf{K}_1\!\wedge\!\mathbf{K}_2\langle\mathbf{S}_1\mathbf{S}_2\rangle_2,\\\mathbf{J}_1\!\wedge\!\mathbf{K}_2\!+\!\mathbf{J}_2\!\wedge\!\mathbf{K}_1=
\mathbf{S}_1\llcorner\mathbf{K}_2\!-\!\mathbf{S}_2\llcorner\mathbf{K}_1\!+\!(\mathbf{K}_1\!+\!\mathbf{J}_1)\omega_2=0=\sigma_2\mathbf{K}_1\gamma_{0123}\!+\!\mathbf{J}_1\!\wedge\!\mathbf{S}_2\!-\!\mathbf{J}_2\!\wedge\!\mathbf{S}_1,\\
\mathbf{S}_1\wedge\mathbf{S}_2=0=\mathbf{J}_1\lrcorner\mathbf{K}_2+\mathbf{J}_2\lrcorner\mathbf{K}_1.
\eeq
\end{subequations}
\end{enumerate}
\item[2)] The fourth term in the brackets, in Eq. (\ref{2ndsigma}), $
a^\dagger_{{\bf p},s}b^\dagger_{{\bf p}',s'}\bar\pwp_1^s(p)\pwpw^{s'}(p')$ 
has $\bar\pwp_1^s(p)\pwpw^{s'}(p')$ as the core spinor content, that is what matters for the analysis of the conditions for the scalar covariant bilinear $\upsigma$ to be zero. 
For this case, the analysis is identical to the one presented in the item 1) above. 
\item[3)] To the second term in the brackets, in Eq. (\ref{2ndsigma}), 
\beq\label{caso3}
b_{{\bf p},s}b^\dagger_{{\bf p}',s'}\bar\pwp_2^s(p)\pwpe^{s'}(p'), \eeq 
the following possibilities arise:
\begin{itemize}

\item[3.1)]  If the spinors $\pwpw, \pwpe$ are regular spinors then it implies that all terms in Eq. (\ref{z1z2}) do not equal zero. Hence, in this case, the first term of Eq. (\ref{2ndsigma}) does not equal zero.

\item[3.2)]  If the spinors $\pwpw, \pwpe$ are both singular spinors, it means that $\sigma_1=\omega_1=0=\sigma_2=\omega_2$.
Hence, Eq. (\ref{caso3}) reads, by the reconstruction theorem, 
\beq
\bar\pwp_2^s(p)\pwpe^{s'}(p')
&=& \bar\xi_2\bar{Z}_2 Z_2\xi_2=\bar\xi_2\bar{Z}_2^2\xi_2=
\bar\xi_2(4\sigma_2 Z_2)\xi_2\,.
\eeq
This last equality follows from Eq. (\ref{nilp}). Since we analyze here 
singular spinors, one has $\sigma_2=0$, implying that $\bar\pwp_2^s(p)\pwpe^{s'}(p')=0.$ 
\end{itemize}
\item[4)] The third term in the brackets, in Eq. (\ref{2ndsigma}), 
\beq\label{caso33}
a^\dagger_{{\bf p},s}a_{{\bf p}',s'}\bar\pwp_1^s(p)\pwpe^{s'}(p'),
\eeq 
can be further analyzed:
\begin{itemize}
\item[4.1)]  If the spinors $\pwpw, \pwpe$ are regular spinors, then it implies that all terms in Eq. (\ref{z1z2}) do not equal zero. Hence, the first term of Eq. (\ref{2ndsigma}) does not equal zero.

\item[4.2)]  If the spinors $\pwpw, \pwpe$ are both singular spinors, it means that $\sigma_1=\omega_1=0=\sigma_2=\omega_2$.
Hence, Eq. (\ref{caso33}) reads, by the reconstruction theorem, 
\beq
\bar\pwp_1^s(p)\pwpw^{s'}(p')
&=& \bar\xi_1\bar{Z}_1 Z_1\xi_1=\bar\xi_1\bar{Z}_1^2\xi_1=
\bar\xi_1(4\sigma_1 Z_1)\xi_1\,.
\eeq
This last equality is due to Eq. (\ref{nilp}). Since for singular spinors, one has $\sigma_1=0$, implying that
$\bar\pwp_2^s(p)\pwpe^{s'}(p')=0$. \end{itemize} \end{enumerate}
    
    To summarize, the vanishing values of the first-quantized scalar and pseudoscalar bilinear covariants do not guarantee 
    that the second quantized scalar bilinear ones shall vanish, too. In order for this to happen, further conditions  studied in details in the above items 1)-2)    should hold. 
    \subsection{pseudoscalar bilinear covariant}
    Now we shall compute the pseudoscalar bilinear covariant for quantum fields
\ba\label{2ndomega}
 \upomega=\wpwp(x)\;\gamma_{0123}\;\pwp(x')&=&\left. \int \frac{d^3{\bf p}d^3{\bf p}'}{(2\pi)^6 2\sqrt{E_{\bf p}E_{{\bf p}'}}} \sum_{s,s'=1,2}\Bigg[b_{{\bf p},s}a_{{\bf p}',s'}\bar\pwp_2^s(p)\gamma_{0123}\pwpw^{s'}\!(p')e^{-i(p\cdot x+p'\cdot x')}\right.\nonumber\\&+&\left.b_{{\bf p},s}b^\dagger_{{\bf p}',s'}\bar\pwp_2^s(p)\gamma_{0123}\pwpe^{s'}\!(p')e^{i(p'\cdot x'-p\cdot x)}+a^\dagger_{{\bf p},s}a_{{\bf p'},s'}\bar\pwp_1^s(p)\gamma_{0123}\pwp_1^{s'}\!(p')e^{i(p\cdot x - p'\cdot x')}\right.\nonumber\\&+&\left.a^\dagger_{{\bf p},s}b^\dagger_{{\bf p}',s'}\bar\pwp_1^s(p)\gamma_{0123}\uppsi_2^{s'}\!(p')e^{i(p\cdot x+p'\cdot x')} \Bigg]\right. . 
\ea Let us analyze the second quantized scalar bilinear covariant $\upomega$ in a similar fashion of what was performed in the last subsection. Eq. (\ref{2ndomega}) has four terms that shall be scrutinized. When all such terms equal zero, then $ \upomega=0$.
   \begin{enumerate}
    \item[I)] In what follows, again, the core spinor part of the term $b_{{\bf p},s}a_{{\bf p}',s'}\bar\pwp_2^s(p)\gamma_{0123}\pwpw^{s'}(p')$ shall be considered, \beq\bar\pwp_2^s(p)\gamma_{0123}\pwpw^{s'}(p').  \label{ppli2}  \eeq 
    The reconstruction theorem yields  %%%%%%%%%%%%%AQUI%%%%%%%%%%%%%%%%%%%
     \beq\bar\pwp_2^s(p)\gamma_{0123}\pwpw^{s'}(p')=    \bar\xi_2^s(p)\bar{Z_2}\gamma_{0123}Z_1\xi_1^{s'}(p').\label{z1z22}\eeq
     Since $\bar{Z_2}\gamma_{0123}Z_1$ is, in general, a multivector,  to analyze whether the term $\bar\pwp_2^s(p)\gamma_{0123}\pwpw^{s'}(p')$ is null resides on the scrutiny of  $\bar{Z_2}\gamma_{0123}Z_1$. By the definition of the boomerang $\bar{Z}_a=Z_a$,    since the homogeneous parts of (\ref{boome}) correspond to the bilinear covariants of some spinor \cite{lou2}. 
 Eq. (\ref{z1z22}) can be then computed, from the aggregates for the spinors $\pwpw$ and $\pwpe$, respectively:
 \begin{subequations}
\beq
{\rm Z}_1(p)&=&\sigma_1(p)+\mathbf{J}_1(p)+i\mathbf{S}_1(p)+(i\mathbf{K}_1(p)+\omega_1(p))\gamma_{0123},\qquad\quad\\
{\rm Z}_2(p')&=&\sigma_2(p')+\mathbf{J}_2(p')+i\mathbf{S}_2(p')+(i\mathbf{K}_2(p')+\omega_2(p'))\gamma_{0123}.    \label{z2}
\eeq
\end{subequations}
Once again the subindex ``1'' [``2''] is associated to bilinears evaluated at the point $p$ [$p'$]. 
Hence, by employing the above equations for the boomerangs for the spinors $\pwpw$ and $\pwpe$,  it yields the following complex multivector:
    \beq\label{z32}
    {\rm Z}_2\gamma_{0123}{\rm Z}_1 &=&(\sigma_1\!+\!\mathbf{J}_1\!+\!i\mathbf{S}_1\!+\!i\mathbf{K}_1\gamma_{0123}\!+\!\omega_1\gamma_{0123})(\sigma_2\gamma_{0123}\!-\!\mathbf{J}_2\gamma_{0123}\!+\!i\mathbf{S}_2\gamma_{0123}\!-\!i\mathbf{K}_2\!+\!\omega_2)\nonumber\\
    &=&(\sigma_1\sigma_2\!-\!\mathbf{J}_1\mathbf{J}_2\!-\!\mathbf{S}_1\mathbf{S}_2\!-\!\mathbf{K}_1\mathbf{K}_2\!+\!\omega_1\omega_2
    \!+\!\sigma_1\mathbf{J}_2-\sigma_2\mathbf{J}_1\!+\!\mathbf{J}_1\omega_2\!+\!\mathbf{J}_2\omega_1)\gamma_{0123}\!+\!\sigma_1\omega_2\!+\!\omega_1\sigma_2
\!+\!\mathbf{S}_1\mathbf{K}_2\!+\!\mathbf{S}_2\mathbf{K}_1
\nonumber\\
&&\!\!\!\!\!\!\!\!+i\left[(\sigma_1(\mathbf{S}_2\!+\!\mathbf{K}_2)\!+\!\sigma_2(\mathbf{S}_1\!+\!\mathbf{K}_1)\!+\!\mathbf{J}_1\mathbf{S}_2\!+\!\mathbf{J}_2\mathbf{S}_1)\gamma_{0123}\!-\!(\mathbf{J}_1\mathbf{K}_2\!+\!\mathbf{J}_2\mathbf{K}_1)\!+\!(\mathbf{S}_1\!+\!\mathbf{K}_1)\omega_2\!+\!(\mathbf{S}_2\!+\!\mathbf{K}_2)\omega_1\right].\eeq The above expression, Eq. (\ref{z32}), shall be now analyzed to verify
all the possibilities to make the term (\ref{ppli2}) to be null.

\item[I.1)] Being the spinors $\pwpw, \pwpe$ are regular, it implies that all terms in Eq. (\ref{z1z22}) do not equal zero. Hence, for regular spinors, the first term of Eq. (\ref{2ndsigma}) does not equal zero.
\item[I.2)] In the case the spinors $\pwpw, \pwpe$ are both singular spinors, hence Eq. (\ref{z32}) reads
 \beq\label{z30}
    {\rm Z}_2\gamma_{0123}{\rm Z}_1     &=&(\mathbf{J}_1\mathbf{J}_2-\mathbf{S}_1\mathbf{S}_2-\mathbf{K}_1\mathbf{K}_2)\gamma_{0123}
+\mathbf{S}_1\mathbf{K}_2+\mathbf{S}_2\mathbf{K}_1
+i\left[(\mathbf{J}_1\mathbf{S}_2\!+\!\mathbf{J}_2\mathbf{S}_1)\gamma_{0123}  -(\mathbf{J}_1\mathbf{K}_2\!+\!\mathbf{J}_2\mathbf{K}_1)\right].\eeq
In order to have $ {\rm Z}_2\gamma_{0123}{\rm Z}_1=0$, both the real and the complex part must equal zero. Hence, we shall 
scrutinize the subcases that follow from this case I.2):
\begin{enumerate}
\item[I.2.1)] For the case where $\pwpw,\pwpe$ are both type-4 one has, according to Eq. (\ref{tipo4}), the values for the bilinear covariants $
\sigma_a=0=\omega_a, \;\;\mathbf{K}_a\neq0\;\;\mathbf{S}_a\neq0\,.$ 
Now, let us consider Eq. (\ref{z00}) and see what are the conditions that make the complex multivector ${\rm Z}_2{\rm Z}_1$ to equal zero. For it,   ${\rm Z}_2{\rm Z}_1$ in Eq. (\ref{z30}) can be split into its non-zero homogeneous parts, 
\begin{subequations}
\beq\label{3ssqq}
&&\!\!\!\!\!\!\!\!\!\!\!\!\!\!\!\!\!\!\mathbf{J}_1\lrcorner\mathbf{J}_2-\mathbf{S}_1\lrcorner\mathbf{S}_2+\mathbf{K}_1\lrcorner\mathbf{K}_2
\in\sec\Omega_{}^4(M),\\
&&\!\!\!\!\!\!\!\!\!\!\!\!\!\!\!\!\!\!(\mathbf{S}_1\wedge\mathbf{K}_2+\mathbf{S}_2\wedge\mathbf{K}_1)\gamma_{0123}\in\sec\Omega_{}^3(M),\\
&&\!\!\!\!\!\!\!\!\!\!\!\!\!\!\!\!\!\!\mathbf{J}_1\!\wedge\!\mathbf{J}_2\!+\!\mathbf{K}_1\!\wedge\!\mathbf{K}_2\!+\!\langle\mathbf{S}_1\mathbf{S}_2\rangle_2
\!-\!i(\mathbf{J}_1\!\wedge\!\mathbf{K}_2\!+\!\mathbf{J}_2\!\wedge\!\mathbf{K}_1)\gamma_{0123}\in\sec\Omega_{}^2(M),\\
&&\!\!\!\!\!\!\!\!\!\!\!\!\!\!\!\!\!\!(\mathbf{S}_1\llcorner\mathbf{K}_2-\mathbf{S}_2\llcorner\mathbf{K}_1)\gamma_{0123}+i(\mathbf{J}_1\wedge\mathbf{S}_2-\mathbf{J}_2\wedge\mathbf{S}_1)\;\;\in\sec\Omega_{}^1(M),\\
&&\!\!\!\!\!\!\!\!\!\!\!\!\!\!\!\!\!\!-\mathbf{S}_1\wedge\mathbf{S}_2-i\left[\mathbf{J}_1\lrcorner\mathbf{K}_2+\mathbf{J}_2\lrcorner\mathbf{K}_1\right]\gamma_{0123}\;\in\sec\Omega_{}^0(M).
\eeq
\end{subequations}
To verify which are the conditions that the bilinear covariants must satisfy to force ${\rm Z}_2{\rm Z}_1$ to be zero,  Eqs. (\ref{3ssqq}) must be equaled to zero, yielding  the following simultaneous conditions:
\begin{subequations}
\beq\label{3cflagdi}
\!\!\!\!\!\!\!\!\!\!\mathbf{J}_1\lrcorner\mathbf{J}_2-\mathbf{S}_1\lrcorner\mathbf{S}_2+\mathbf{K}_1\lrcorner\mathbf{K}_2
=0=
\mathbf{S}_1\wedge\mathbf{K}_2+\mathbf{S}_2\wedge\mathbf{K}_1=
\mathbf{J}_1\!\wedge\!\mathbf{J}_2\!+\!\mathbf{K}_1\!\wedge\!\mathbf{K}_2\!+\!\langle\mathbf{S}_1\mathbf{S}_2\rangle_2\!=\!\mathbf{J}_1\!\wedge\!\mathbf{K}_2\!+\!\mathbf{J}_2\!\wedge\!\mathbf{K}_1,\\
\!\!\!\!\!\mathbf{S}_1\wedge\mathbf{S}_2=0=\mathbf{J}_1\lrcorner\mathbf{K}_2+\mathbf{J}_2\lrcorner\mathbf{K}_1\,.
\label{3418}
\eeq
\end{subequations}
When Eqs. (\ref{3cflagdi}) and (\ref{3418}) hold, it means that the second quantized scalar bilinear covariant is null.

\item[I.2.2)] For the case where $\pwpw,\pwpe$ are both type-5, let us consider Eq. (\ref{z30}) and see what are the conditions that make the complex multivector ${\rm Z}_2{\rm Z}_1$ to equal zero. For it, let us split ${\rm Z}_2{\rm Z}_1$, Eq. (\ref{z30}),  into its non-zero homogeneous parts, 
\begin{subequations}
\beq\label{3ssq1}
&&\!\!\!\!\!\!\!\!\!\!\!\!\!\!\!\!\!\!\mathbf{J}_1\lrcorner\mathbf{J}_2-\mathbf{S}_1\lrcorner\mathbf{S}_2\in\sec\Omega_{}^4(M),\\
&&\!\!\!\!\!\!\!\!\!\!\!\!\!\!\!\!\!\!\mathbf{J}_1\!\wedge\!\mathbf{J}_2\!+\!\langle\mathbf{S}_1\mathbf{S}_2\rangle_2
\in\sec\Omega_{}^2(M),\\
&&\!\!\!\!\!\!\!\!\!\!\!\!\!\!\!\!\!\!i(\mathbf{J}_1\wedge\mathbf{S}_2-\mathbf{J}_2\wedge\mathbf{S}_1)\;\;\in\sec\Omega_{}^1(M),\\
&&\!\!\!\!\!\!\!\!\!\!\!\!\!\!\!\!\!\!-\mathbf{S}_1\wedge\mathbf{S}_2\gamma_{0123}\;\in\sec\Omega_{}^0(M).\label{3ssq11}
\eeq
\end{subequations}
As usual, to verify the conditions yielding  ${\rm Z}_2{\rm Z}_1=0$ Eqs. (\ref{3ssq1}) - (\ref{3ssq11}) must vanish or, equivalently, \beq\label{3cflag}
\mathbf{J}_1\lrcorner\mathbf{J}_2-\mathbf{S}_1\lrcorner\mathbf{S}_2
=0=\mathbf{J}_1\!\wedge\!\mathbf{J}_2\!+\!\langle\mathbf{S}_1\mathbf{S}_2\rangle_2=\mathbf{S}_1\wedge\mathbf{S}_2\,.
\label{3419}
\eeq
When Eqs. (\ref{3419}) hold, it means that the second quantized scalar bilinear covariant is null.

\item[I.2.3)] For the case where $\pwpw,\pwpe$ are both type-6, the conditions that make the complex multivector ${\rm Z}_2{\rm Z}_1$ to equal zero can be obtained by first splitting ${\rm Z}_2{\rm Z}_1$ in Eq. (\ref{z30}) in its non-zero homogeneous parts,\begin{subequations}
\beq\label{3ssq6}
&&\!\!\!\!\!\!\!\!\!\!\!\!\!\!\!\!\!\!\mathbf{J}_1\lrcorner\mathbf{J}_2+\mathbf{K}_1\lrcorner\mathbf{K}_2
\in\sec\Omega_{}^0(M),\\
&&\!\!\!\!\!\!\!\!\!\!\!\!\!\!\!\!\!\!\mathbf{J}_1\!\wedge\!\mathbf{J}_2\!+\!\mathbf{K}_1\!\wedge\!\mathbf{K}_2\!-\!i(\mathbf{J}_1\!\wedge\!\mathbf{K}_2\!+\!\mathbf{J}_2\!\wedge\!\mathbf{K}_1)\gamma_{0123}\in\sec\Omega_{}^2(M),\\
&&\!\!\!\!\!\!\!\!\!\!\!\!\!\!\!\!\!\!-i\left[\mathbf{J}_1\lrcorner\mathbf{K}_2+\mathbf{J}_2\lrcorner\mathbf{K}_1\right]\gamma_{0123}\;\in\sec\Omega_{}^4(M).\label{3ssq61}
\eeq
\end{subequations}
For ${\rm Z}_2{\rm Z}_1$ to be zero,  Eqs. (\ref{3ssq6})-(\ref{3ssq61}) must be equal to zero, yielding the following simultaneous conditions:
\beq\label{3cflagdip}
\mathbf{J}_1\lrcorner\mathbf{J}_2+\mathbf{K}_1\lrcorner\mathbf{K}_2
=0=
\mathbf{J}_1\!\wedge\!\mathbf{J}_2\!+\!\mathbf{K}_1\!\wedge\!\mathbf{K}_2=\mathbf{J}_1\!\wedge\!\mathbf{K}_2\!+\!\mathbf{J}_2\!\wedge\!\mathbf{K}_1=
\mathbf{J}_1\lrcorner\mathbf{K}_2+\mathbf{J}_2\lrcorner\mathbf{K}_1\,,
\label{3420}
\eeq
also yielding the second quantized scalar bilinear covariant to be null, 
\beq
\upomega = \;\wpwp\;\gamma_{0123}\;\pwp\,=0\,.
\eeq

\item[I.2.4)] For the case where $\pwp_2$ is a regular spinor, of type-1, and $\pwp_1$ is a type-5 spinor (noticed that these roles are interchangeable, $\pwpw\leftrightarrow\pwpe$), and splitting Eq. (\ref{z30}) into its homogeneous 
parts yields
\begin{subequations}
\beq\label{31ssqq}
&&\!\!\!\!\!\!\!\!\!\!\!\!\!\!\!\!\!\!\mathbf{J}_1\lrcorner\mathbf{J}_2-\mathbf{S}_1\lrcorner\mathbf{S}_2
\in\sec\Omega_{}^4(M),\\
&&\!\!\!\!\!\!\!\!\!\!\!\!\!\!\!\!\!\!\sigma_2\mathbf{J}_1+\mathbf{S}_1\wedge\mathbf{K}_2\gamma_{0123}\in\sec\Omega_{}^3(M),\\
&&\!\!\!\!\!\!\!\!\!\!\!\!\!\!\!\!\!\!\mathbf{J}_1\!\wedge\!\mathbf{J}_2\!+\!\langle\mathbf{S}_1\mathbf{S}_2\rangle_2\!+\mathbf{S}_1\omega_2\gamma_{0123}\!
\!-\!i(\sigma_2\mathbf{S}_1+\mathbf{J}_1\!\wedge\!\mathbf{K}_2\gamma_{0123})\in\sec\Omega_{}^2(M),\\
&&\!\!\!\!\!\!\!\!\!\!\!\!\!\!\!\!\!\!(\mathbf{J}_1\omega_2+\mathbf{S}_1\llcorner\mathbf{K}_2)\gamma_{0123}+i(\mathbf{J}_1\wedge\mathbf{S}_2-\mathbf{J}_2\wedge\mathbf{S}_1)\;\;\in\sec\Omega_{}^1(M),\\
&&\!\!\!\!\!\!\!\!\!\!\!\!\!\!\!\!\!\!-\mathbf{S}_1\wedge\mathbf{S}_2-i\mathbf{J}_1\lrcorner\mathbf{K}_2\gamma_{0123}\;\in\sec\Omega_{}^0(M).
\eeq
\end{subequations} The reader is certainly evincing the similarity with the cases $1.2.3, 1.2.4$ and so on of the previous subsection. We call attention to the difference sometimes explicit in the section of the exterior bundle. Returning to our analysis, for ${\rm Z}_2{\rm Z}_1=0$, the following equations must hold, simultaneously:
\begin{subequations}
\beq\label{33ssqq}
\!\!\!\!\!\!\!\!\!\!\!\!\!\!\!\!\!\!\!\!\!\!\!\!\!\!\!\!\mathbf{J}_1\lrcorner\mathbf{J}_2-\mathbf{S}_1\lrcorner\mathbf{S}_2=0=
\sigma_2\mathbf{J}_1+\mathbf{S}_1\wedge\mathbf{K}_2\gamma_{0123}=
\mathbf{J}_1\!\wedge\!\mathbf{J}_2\!+\!\langle\mathbf{S}_1\mathbf{S}_2\rangle_2\!+\mathbf{S}_1\omega_2\gamma_{0123}=\sigma_2\mathbf{S}_1+\mathbf{J}_1\!\wedge\!\mathbf{K}_2\gamma_{0123},\\
(\mathbf{J}_1\omega_2+\mathbf{S}_1\llcorner\mathbf{K}_2)\gamma_{0123}=0=\mathbf{J}_1\wedge\mathbf{S}_2-\mathbf{J}_2\wedge\mathbf{S}_1=
\mathbf{S}_1\wedge\mathbf{S}_2=\mathbf{J}_1\lrcorner\mathbf{K}_2\gamma_{0123}.
\eeq
\end{subequations}

\item[I.2.5)] For the case where $\pwpe$ is a regular spinor and $\pwpw$ is a type-6 one has, according to Eq. (\ref{tipo6}), the values for the bilinear covariants:
\beq
\sigma_1=0=\omega_1, \;\;\mathbf{S}_1=0.
\eeq
By splitting Eq. (\ref{z30}) into its homogeneous 
parts yields
\begin{subequations}
\beq\label{3ssq66}
&&\!\!\!\!\!\!\!\!\!\!\!\!\!\!\!\!\!\!\!\!\!\!\mathbf{J}_1\lrcorner\mathbf{J}_2+\mathbf{K}_1\lrcorner\mathbf{K}_2
\in\sec\Omega_{}^4(M),\\
&&\!\!\!\!\!\!\!\!\!\!\!\!\!\!\!\!\!\!\!\!\!\!\sigma_2\mathbf{J}_1+\mathbf{S}_2\wedge\mathbf{K}_1\gamma_{0123}\in\sec\Omega_{}^3(M),\\
&&\!\!\!\!\!\!\!\!\!\!\!\!\!\!\!\!\!\!\!\!\!\!\mathbf{J}_1\!\wedge\!\mathbf{J}_2\!+\!\mathbf{K}_1\!\wedge\!\mathbf{K}_2\!-\!i(\mathbf{J}_1\!\wedge\!\mathbf{K}_2\!+\!\mathbf{J}_2\!\wedge\!\mathbf{K}_1)\gamma_{0123}\in\sec\Omega_{}^2(M),\\
&&\!\!\!\!\!\!\!\!\!\!\!\!\!\!\!\!\!\!\!\!\!\!\star(-\!\mathbf{S}_2\llcorner\mathbf{K}_1\!+\!(\mathbf{K}_1\!+\!\mathbf{J}_1)\omega_2)\!+\!i(\sigma_2\mathbf{K}_1\gamma_{0123}\!+\!\mathbf{J}_1\wedge\mathbf{S}_2)\in\sec\Omega_{}^1(M),\\
&&\!\!\!\!\!\!\!\!\!\!\!\!\!\!\!\!\!\!\!\!\!\!-\mathbf{S}_1\wedge\mathbf{S}_2-i\left[\mathbf{J}_1\lrcorner\mathbf{K}_2+\mathbf{J}_2\lrcorner\mathbf{K}_1\right]\gamma_{0123}\;\in\sec\Omega_{}^0(M).
\eeq
\end{subequations}
Hence, for ${\rm Z}_2{\rm Z}_1=0$ to hold, the following equations 
\begin{subequations}
\beq\label{3q66}
\!\!\!\!\!\!\mathbf{J}_1\lrcorner\mathbf{J}_2+\mathbf{K}_1\lrcorner\mathbf{K}_2=0=\sigma_2\mathbf{J}_1+\mathbf{S}_2\wedge\mathbf{K}_1\gamma_{0123}=
\mathbf{J}_1\!\wedge\!\mathbf{J}_2\!+\!\mathbf{K}_1\!\wedge\!\mathbf{K}_2=\mathbf{J}_1\!\wedge\!\mathbf{K}_2\!+\!\mathbf{J}_2\!\wedge\!\mathbf{K}_1,\\
-\!\mathbf{S}_2\llcorner\mathbf{K}_1\!+\!(\mathbf{K}_1\!+\!\mathbf{J}_1)\omega_2\!\!=0=\!\sigma_2\mathbf{K}_1\gamma_{0123}\!+\!\mathbf{J}_1\!\wedge\!\mathbf{S}_2\!=
\mathbf{J}_1\lrcorner\mathbf{K}_2+\mathbf{J}_2\lrcorner\mathbf{K}_1,
\eeq\end{subequations} must be satisfied.

\item[I.2.6)] For the case where $\pwpw$ is a regular spinor and $\pwpe$ is a type-4 spinor one has, according to Eq. (\ref{tipo4}), the following splitting of ${\rm Z}_2{\rm Z}_1$:
\begin{subequations}
\beq\label{ssq67}
&&\!\!\!\!\!\!\!\!\!\!\!\!\!\!\!\!\!\!\!\!\!\!\mathbf{J}_1\lrcorner\mathbf{J}_2-\mathbf{S}_1\lrcorner\mathbf{S}_2+\mathbf{K}_1\lrcorner\mathbf{K}_2
\in\sec\Omega_{}^4(M),\\
&&\!\!\!\!\!\!\!\!\!\!\!\!\!\!\!\!\!\!\!\!\!\!\sigma_2\mathbf{J}_1+(\mathbf{S}_1\wedge\mathbf{K}_2+\mathbf{S}_2\wedge\mathbf{K}_1)\gamma_{0123}\in\sec\Omega_{}^3(M),\\
&&\!\!\!\!\!\!\!\!\!\!\!\!\!\!\!\!\!\!\!\!\!\!\mathbf{J}_1\!\wedge\!\mathbf{J}_2\!+\!\mathbf{K}_1\!\wedge\!\mathbf{K}_2\!+\!\langle\mathbf{S}_1\mathbf{S}_2\rangle_2
\!-\!i(\mathbf{J}_1\!\wedge\!\mathbf{K}_2\!+\!\mathbf{J}_2\!\wedge\!\mathbf{K}_1)\gamma_{0123}\in\sec\Omega_{}^2(M),\\
&&\!\!\!\!\!\!\!\!\!\!\!\!\!\!\!\!\!\!\!\!\!\!\star(\mathbf{S}_1\llcorner\mathbf{K}_2\!-\!\mathbf{S}_2\llcorner\mathbf{K}_1\!+\!(\mathbf{K}_1\!+\!\mathbf{J}_1)\omega_2)\!+\!i(\sigma_2\mathbf{K}_1\gamma_{0123}\!+\!\mathbf{J}_1\wedge\mathbf{S}_2\!-\!\mathbf{J}_2\wedge\mathbf{S}_1)\in\sec\Omega_{}^1(M),\\
&&\!\!\!\!\!\!\!\!\!\!\!\!\!\!\!\!\!\!\!\!\!\!-\mathbf{S}_1\wedge\mathbf{S}_2-i\left[\mathbf{J}_1\lrcorner\mathbf{K}_2+\mathbf{J}_2\lrcorner\mathbf{K}_1\right]\gamma_{0123}\;\in\sec\Omega_{}^0(M).
\eeq
\end{subequations}
Hence, for ${\rm Z}_2{\rm Z}_1=0$, the following equations must simultaneously hold: 
\begin{subequations}
\beq\label{q66}
\hspace*{-2cm}\!\!\!\!\!\!\!\!\!\!\!\!\!\!\!\!\!\!\!\!\!\!\!\!\!\!\!\!\!\!\!\!\!\!\!\!\!\!\!\!\!\!\!\!\!\!\!\!\!\!\!\!\!\!\!\!\!\!\!\!\!\!\!\!\!\!\mathbf{J}_1\lrcorner\mathbf{J}_2\!-\!\mathbf{S}_1\lrcorner\mathbf{S}_2\!+\!\mathbf{K}_1\lrcorner\mathbf{K}_2\!=\!0\!=\!
\sigma_2\mathbf{J}_1\!+\!(\mathbf{S}_1\!\wedge\!\mathbf{K}_2\!+\!\mathbf{S}_2\!\wedge\!\mathbf{K}_1)\gamma_{0123}\!=\!
\mathbf{J}_1\!\wedge\!\mathbf{J}_2\!\!+\!\!\mathbf{K}_1\!\wedge\!\mathbf{K}_2\!\!+\!\!\langle\mathbf{S}_1\mathbf{S}_2\rangle_2,\\\!\!\!\!\!\!\!\!\!\!\!\!\!\!\!\!\!\!\!\!\!\!\mathbf{J}_1\!\wedge\!\mathbf{K}_2\!\!+\!\!\mathbf{J}_2\!\wedge\!\mathbf{K}_1\!=\!0\!=\!
\mathbf{S}_1\llcorner\mathbf{K}_2\!\!-\!\!\mathbf{S}_2\llcorner\mathbf{K}_1\!\!+\!\!(\mathbf{K}_1\!\!+\!\!\mathbf{J}_1)\omega_2\!=\!\sigma_2\mathbf{K}_1\gamma_{0123}\!\!+\!\!\mathbf{J}_1\!\wedge\!\mathbf{S}_2\!-\!\mathbf{J}_2\!\wedge\!\mathbf{S}_1\!=\!
\mathbf{S}_1\wedge\mathbf{S}_2\!=\!\mathbf{J}_1\lrcorner\mathbf{K}_2\!+\!\mathbf{J}_2\lrcorner\mathbf{K}_1.
\eeq
\end{subequations}

\end{enumerate}
\item[II)] 
The fourth term in the brackets, in Eq. (\ref{2ndomega}), $
a^\dagger_{{\bf p},s}b^\dagger_{{\bf p}',s'}\bar\pwp_1^s(p)\gamma_{0123}\pwpw^{s'}(p')$ 
has
$\bar\pwp_1^s(p)\gamma_{0123}\pwpw^{s'}(p')$ as the core spinor content, that is what matters for the analysis of the conditions for the scalar covariant bilinear $\upomega$ to be zero.
For this case, the analysis is identical to the one in the item I) above. 
\item[III)] The second term in brackets, in Eq. (\ref{2ndomega}), 
\beq\label{caso3pse}
b_{{\bf p},s}b^\dagger_{{\bf p}',s'}\bar\pwp_2^s(p)\gamma_{0123}\pwpe^{s'}(p'), 
\eeq has a core spinor content that must be further analyzed:
\begin{itemize}
\item[III.1)]  If the spinors $\pwpw, \pwpe$ are regular, then it implies that all terms in Eq. (\ref{z1z22}) do not equal zero. Hence, for regular spinors, the first term of Eq. (\ref{2ndomega}) does not equal zero.

\item[III.2)]  If the spinors $\pwpw, \pwpe$ are both singular, it means that $\sigma_1=\omega_1=0=\sigma_2=\omega_2$.
Hence, Eq. (\ref{caso3pse}) reads, by the reconstruction theorem, 
\beq
\bar\pwp_2^s(p)\gamma_{0123}\pwpe^{s'}(p')
&=& \bar\xi_2\bar{Z}_2 \gamma_{0123}Z_2\xi_2=\xi_2^\dagger{Z}_2 \gamma_{0123}Z_2\xi_2=
\xi_2^\dagger(-4\omega_2 Z_2)\xi_2\,.
\eeq
This last equality is due to Eq. (\ref{nilp}). Since we analyze here 
singular spinors, one has $\sigma_2=0$, implying that $\bar\pwp_2^s(p)\gamma_{0123}\pwpe^{s'}(p')=0.$ 
\end{itemize}
\item[IV)] Now, the third term in the brackets in Eq. (\ref{2ndsigma}), $
a^\dagger_{{\bf p},s}a_{{\bf p}',s'}\bar\pwp_1^s(p)\gamma_{0123}\pwpe^{s'}(p'),$ 
has
\beq\label{caso3star}
\bar\pwp_2^s(p)\gamma_{0123}\pwpw^{s'}(p')\eeq as the core spinor content.
\begin{itemize}
\item[IV.1)]  If the spinors $\pwpw, \pwpe$ are regular,  then it implies that all terms in Eq. (\ref{z1z22}) do not equal zero. Hence, for regular spinors, the first term of Eq. (\ref{2ndsigma}) does not equal zero.

\item[IV.2)]  If the spinors $\pwpw, \pwpe$ are both singular, it means that $\sigma_1=\omega_1=0=\sigma_2=\omega_2$. Hence, Eq. (\ref{caso3star}) reads, by the reconstruction theorem, 
\beq
\bar\pwp_1^s(p)\gamma_{0123}\pwpw^{s'}(p')
&=& \bar\xi_1^\dagger\bar{Z}_1\gamma_{0123} Z_1\xi_1=\xi_1^\dagger{Z}_1\gamma_{0123} Z_1\xi_1=
\xi_1(-4\omega_1 Z_1)\xi_1\,.
\eeq
This last equality is due to Eq. (\ref{nilp}). Since we analyze here singular spinors, one has $\sigma_1=0$, implying that
\beq
\bar\pwp_2^s(p)\gamma_{0123}\pwpe^{s'}(p')=0
.\eeq\end{itemize}  \end{enumerate}

      To summarize, the vanishing values of the first-quantized  pseudoscalar bilinear covariants do not guarantee 
    that the second quantized pseudoscalar bilinear  shall further vanish. In order for this to be accomplished, further conditions  studied in details in the above items I)-II)  must hold.

\section{Propagators and Feynman rules}
Heretofore, no assertion concerning the spinor fields dynamics is considered, except the straightforward and exhaustive
fact that all spinors must satisfy the Klein-Gordon equation. Analyzing  spinor fields that satisfy the Dirac equation does not necessarily bring any information on which class this spinor field does belong to in the Lounesto's classification. In fact, although 
solutions of the Dirac equation have been found in the literature to occupy all the Lounesto's spinor classes \cite{esk,daRocha:2016bil,daSilva:2012wp}, they are  far from encompassing all the spinors in each spinor class. Besides the Weyl, Majorana and 
Elko spinors, there are more types of spinors with unknown dynamics in the Lounesto's classification.

Hence, in this section we extend the calculations of $n$-point functions and propagators to all the spinors in the Lounesto's classification that satisfy the Dirac equation, as well as for eigenspinors of the charge conjugation operator that have mass dimension $3/2$ in Minkowski spacetime, i. e., Majorana spinors. 

Remember that for a Dirac field $\psi(x)$, one has 
\begin{subequations}
    \beq
    \br\,\psi(x)\,\ke &=& 0,\label{lor1}\\
    \bra{p,s,+\,}\,\psi(x)\,\ke&=&0=  \bra{p,s,-\,}\,\bar\psi(x)\,  \ke,\label{cco1}\\
    \bra{p,s,-\,}\,\psi(x)\,\ke&=&v_s(p)e^{-ipx},\label{lla1}\\
    \bra{p,s,+\,}\,\bar\psi(x)\,\ke&=&\bar{u}_s(p)e^{-ipx}.\label{lla2}         \eeq\end{subequations}
    Eqs. (\ref{cco1}) follow from the charge conservation, whereas Eq. (\ref{lor1}) is required by covariance. 
    For neutral fields, Eq. (\ref{lor1}) holds still, as well as the analogue of  Eqs. (\ref{lla1}, \ref{lla2}):
     \beq
     \bra{p,s}\,\psi(x)\,\ke=v_s(p)e^{-ipx},\label{lla3} \qquad 
    \bra{p,s}\,\bar\psi(x)\,\ke=\bar{u}_s(p)e^{-ipx},
           \eeq\noindent for uncharged fields, where 
     \begin{subequations}  \beq\label{psi11}
 \psi(x)&=&\int \frac{d^3{\bf p}}{(2\pi)^3\sqrt{2E_{\bf p}}}\sum_{s=1,2}\;\left(a_{{\bf p},s} u^s(p)\,e^{-ip\cdot{x}}+a^\dagger_{{\bf p},s} v^s(p)\,e^{i{p}\cdot{x}}\right),\\\label{barpsi11}
 \bar\psi(x)&=&\int \frac{d^3{\bf p}}{(2\pi)^3\sqrt{2E_{\bf p}}}\sum_{s=1,2}\;\left(a^\dagger_{{\bf p},s} \bar{u}^s(p)\,e^{ip\cdot{x}}+a_{{\bf p},s} \bar{v}^s(p)\,e^{-i{p}\cdot{x}}\right).\eeq\end{subequations}

 Now, assuming any kind of spinor satisfying Eq. (\ref{2ndsigma}),
 one can show that the Feynman propagator constructed upon these quantum fields is the same as for the Dirac fermion. {\color{black}{In fact, the textbook Dirac spinors are eigenspinors of the parity operator that reside in the class 1, Eq. (\ref{tipo1}), of Lounesto's classification \cite{lou2}, and there are also regular fermions in classes 2 and 3, respectively given by Eqs. (\ref{tipo2}, \ref{tipo3}), that satisfy the Dirac equation \cite{Vaz:2016qyw}. Besides, 
 Ref. \cite{esk} showed that a type-4 flag-dipole in the class (\ref{tipo4}) of singular spinor also satisfy the Dirac equation, as well as a peculiar, recent found, type-5 flagpole spinor \cite{daRocha:2016bil}. Although 
 the dynamics in each class of Lounesto's classification is an open issue, at least the specific spinors above described have spinors constituting the respective fermion quantum fields  that satisfy the Dirac equation. Therefore, the  Feynman propagators, constructed upon these quantum fields, are analogous to  the standard propagator for the Dirac fermion.}}

Now, given the Majorana condition $\bar\uppsi = \uppsi^\intercal C$, where $C$ denotes the charge conjugation operator and $(\;\;)^\intercal$ denotes the real adjoint operator, imposes that the correlators $ \langle \;0\,|\,T(\bar\uppsi_a(x) \uppsi_b(y))\,|\,0\,\rangle$ and $ \langle \;0\,|\,T(\bar\uppsi_a(x) \uppsi_b(y))\,|\,0\,\rangle$ do not vanish. In fact, 
\bea
 \langle \;0\,|\,T(\uppsi_a(x) \uppsi_b(y))\,|\,0\,\rangle =  \langle \;0\,|\,T(\uppsi_a(x) \bar\uppsi_c(y))\,|\,0\,\rangle (C^{-1})_{cb}= [S_{}(x-y)C^{-1}]_{ab}.
 \eea
 where $S_{ab}(x-y)$ again denotes the standard Dirac propagator.
 A similar proof \textcolor{black}{yields} another 2-point function:
 \bea
 \langle \;0\,|\,T(\bar\uppsi_a(x) \bar\uppsi_b(y))\,|\,0\,\rangle =  (C^{-1})_{ac}\langle \;0\,|\,T(\uppsi_c(x) \bar\uppsi_b(y))\,|\,0\,\rangle = [C^{-1}S_{}(x-y)]_{ab}.
 \eea
 Besides, the correlation function of more than two fields \textcolor{black}{in free theories}  read, for Dirac fields, 
 \bea
\!\!\!\!\!\!\!\!\!\!\!\!\!\! \langle \;0\,|\,T(\uppsi_a(x) \bar\uppsi_b(y)\uppsi_c(z) \bar\uppsi_d(w))\,|\,0\,\rangle =  +S_{ab}(x-y)S(z-w)_{cd}-S_{ad}(x-w)S_{cb}(z-y),
 \eea whereas for Majorana quantum fields it reads
\bea
 \langle \,0\,|\,T(\uppsi_a(x) \bar\uppsi_b(y)\uppsi_c(z) \bar\uppsi_d(w))\,|\,0\,\rangle &=&  [S(x-y)C^{-1}]_{ab} [S(z-w)C^{-1}]_{cd}-[S(x-z)C^{-1}]_{ac} [S(y-w)C^{-1}]_{bd}\nonumber\\&&\qquad+[S(x-w)C^{-1}]_{ad} [S(y-z)C^{-1}]_{bd}. \eea

For further reference, recall that for real scalar fields driven by the Lagrangian
    \beq
    \mathcal{L}_0 = -\frac12\partial^\mu\varphi\partial_\mu\varphi-\frac12 m^2\varphi^2=-\frac12\varphi(-\partial^2+m^2)\varphi - \frac12\partial_\mu(\varphi\partial^\mu\varphi),
    \eeq one has the correlation function
   $
    \br\,{\rm T}(\varphi(x_1)\ldots\,)\ke = \frac1i \frac{\delta}{\delta J(x_1)}\ldots Z_0(J)\vert_{J=0},
    $
    where $Z_0(J) = \int\mathcal{D}\varphi\,\exp\left[i\int d^4x(\mathcal{L}_0+J\varphi)\right]=\exp\left[\frac{i}{2}\int d^4xd^4y J(x)\Delta(x-y) J(y)\right]$, for \beq
       \Delta(x-y)= \frac{d^4p}{(2\pi)^4}\frac{e^{ip(x-y)}}{p^2+m^2-i\epsilon},\eeq and $(\partial_x^2+m^2)\Delta(x-y)=\delta^4(x-y)$. For complex scalar fields the results are quite similar. 
%    \beq
%    \mathcal{L}_0 = -\frac12\partial^\mu\varphi^\dagger\partial_\mu\varphi-m^2\varphi^\dagger\varphi=-\varphi(-\partial^2+m^2)\varphi - \partial_\mu(\varphi^\dagger\partial^\mu\varphi). 
%    \eeq 
The correlation function reads 
    \beq
    \br\,{\rm T}(\varphi(x_1)\ldots\varphi^\dagger(y_1)\ldots\,)\ke = \frac1i \frac{\delta}{\delta J^\dagger(x_1)}\ldots \frac1i \frac{\delta}{\delta J(y_1)}\ldots Z_0(J^\dagger, J)\vert_{J=J^\dagger=0},
    \eeq
    where \beq
    Z_0(J^\dagger, J) &=& \int\mathcal{D}\varphi^\dagger\mathcal{D}\varphi\,\exp\left[i\int d^4x(\mathcal{L}_0+J^\dagger\varphi+\varphi^\dagger J)\right]=\exp\left[\frac{i}{2}\int d^4xd^4y J^\dagger(x)\Delta(x-y) J(y)\right].\eeq
%-------------------------------------
    
Returning to our main point, as it is well known, functional derivatives for anti-commuting source variables can be defined as 
    \beq
    \frac{\delta}{\delta \eta(x)}\int d^4y\left[\bar\eta(y)\uppsi(y) + \bar\uppsi(y)\eta(y)\right]=-\bar\uppsi(x),\quad \qquad 
       \frac{\delta}{\delta \bar\eta(x)}\int d^4y\left[\bar\eta(y)\uppsi(y) + \bar\uppsi(y)\eta(y)\right]=+\uppsi(x).\eeq
Now consider, for example, a Yukawa-like theory with a real scalar field interacting with a Dirac field, $
\mathcal{L}_1=g\varphi\bar\uppsi\uppsi,$ 
whose generating functional is
\beq
Z(\bar\eta, \eta, J) \propto \exp\left[ig \int d^4x\left(\frac{1}{i} \frac{\delta}{\delta J^\dagger(x)}\right)\left(i \frac{\delta}{\delta \eta_\alpha(x)}\right)\left(\frac{1}{i} \frac{\delta}{\delta \bar\eta_\alpha(x)}\right)\right]\ldots Z_0(\bar\eta, \eta, J),\label{phipsipsi}
\eeq
where $Z_0(\bar\eta, \eta, J) =\exp\left[i\int d^4xd^4y \bar\eta(x)S(x-y) \eta(y)+\frac{i}{2}\int d^4xd^4y J(x)\Delta(x-y) J(y)\right]$.
Hence,  the term (\ref{phipsipsi}) in the expansion that has one vertex, two fermion
propagators and one scalar propagators reads
    \beq
   && \left[\frac{i}{2}\int d^4xd^4y J(x)\Delta(x-y) J(y)\right]\frac12\left[i\int d^4xd^4y \bar\eta(x)S(x-y) \eta(y)\right]\times\left[i\int d^4xd^4y \bar\eta(x)S(x-y) \eta(y)\right].
    \eeq
    For example, the diagram for the reaction 
    $
    e^+e^- \to \varphi\varphi$ can be computed by calculating the connected correlation function $
    \br\,T(\uppsi\bar\uppsi \varphi\varphi )\ke$, 
    starting with 
    \beq
    \br\,{\rm T}(\uppsi_\alpha(x)\bar\uppsi_\beta(y)\varphi(z_1)\varphi(z_2))\ke&=&\frac1i\frac{\delta}{\delta\bar\eta_\alpha(x)}i\frac{\delta}{\delta\bar\eta_\beta(y)}\frac{1}{i}\frac{\delta}{\delta J(z_1)}\frac{1}{i}\frac{\delta}{\delta J(z_2)}iW(\bar\eta,\eta,J)\vert_{\bar\eta=\eta=J=0}\nonumber\\
    =(-i)^5(ig)^2\!\!\!&&\!\!\!\!\!\!\!\!\!\!\!\int\! 
    d^4w_1 d^4w_2 [S(x\!-\!w_2)S(w_2\!-\!w_1)S(w_1\!-\!y)]_{\alpha\beta}\times\Delta(z_1-w_1)\Delta(z_2-w_2)\nonumber\\\!\!\!\!\!\!\!+(-i)^5(ig)^2\!\!\!&&\!\!\!\!\!\!\!\!\!\!\!\int\! 
    d^4w_1 d^4w_2 [S(x\!-\!w_2)S(w_2\!-\!w_1)S(w_1\!-\!y)]_{\alpha\beta}\Delta(z_2\!-\!w_1)\Delta(z_1\!-\!w_2)\!+\!\mathcal{O}(g^4),
    \eeq and quite similarly
   
      \beq
    \br\,{\rm T}(\uppsi_{\alpha_1}(x_1)\bar\uppsi_{\beta_1}(y_1)\uppsi_{\alpha_2}(x_2)\bar\uppsi_{\beta_2}(y_2))\ke&=&{\scriptsize{\frac1i\frac{\delta}{\delta\bar\eta_{\alpha_1}(x_1)}i\frac{\delta}{\delta\bar\eta_{\beta_1}(y_1)}\frac1i\frac{\delta}{\delta \bar\eta_{\alpha_2}(x_2)}\frac1i\frac{\delta}{\delta \eta_{\beta_2}(y_2)}iW(\bar\eta,\eta,J)\vert_{\bar\eta=\eta=J=0}}}\nonumber\\
    =(-i)^5(ig)^2\!\!\!\int\!\! 
    d^4w_1 d^4w_2&&\!\!\!\!\!\!\!\!\!\! [S(x_1-w_1)S(w_1-y_1)]_{\alpha_1\beta_1}\Delta(w_1\!-\!w_2)[S(x_2\!-w\!_2)S(w_2\!-\!y_2)\nonumber\\\!\!\!\!\!\!\!\!\!\!-(-i)^5(ig)^2\!\!\!\int\!\! 
    d^4w_1 d^4w_2\!\! &\!\!&\!\!\!\!\!\!\!\!\!\![S(x_1\!-\!w_1)S(w_1\!-\!y_2)]_{\alpha_1\beta_2}\Delta(w_1\!-\!w_2)[S(x_2\!-\!w_2)S(w_2\!-\!y_1)
        \!\!+\!\!\mathcal{O}(g^4).
    \eeq

\section{Normal ordered bilinear covariants}

In this section the normal ordered product of the second quantized bilinear covariants shall be performed. In fact, some of them are useful 
to calculate the propagators, when their action on the vacuum of the theory must be taken into account. However, for the general form of the second quantized version of the reconstruction theorem, the normal ordering  is not necessary. 

The general expansion of a quantum field was introduced in Eqs. (\ref{psigen}) and (\ref{psigendual}). Let us then calculate
\beq
\frac12[\bar\uppsi,\gamma^\mu\uppsi]&=&\frac{1}{2(2\pi)^6}\int \frac{d^3{\bf p}d^3{\bf p'}}{(2\pi)^3\sqrt{2E_{\bf p}2E_{\bf p'}}}\sum_{r,s=1,2}\;\Bigg\{
\bar\uppsi_1^r(p)\gamma^\mu \uppsi_1^s(q)(a^\dagger_{{\bf p},r}a_{{\bf q},s}-a_{{\bf q},s}a^\dagger_{{\bf p},r})e^{i(p-q)\cdot x}\nonumber\\
&&\!\!\!\!+\bar\uppsi_1^r(p)\gamma^\mu \uppsi_2^s(q)(a^\dagger_{{\bf p},r}b^\dagger_{{\bf q},s}-b^\dagger_{{\bf q},s}a^\dagger_{{\bf p},r})e^{i(p+q)\cdot x}+\bar\uppsi_2^r(p)\gamma^\mu \uppsi_1^s(q)(b_{{\bf p},r}a_{{\bf q},s}-a_{{\bf q},s}b_{{\bf p},r})e^{-i(p+q)\cdot x}\nonumber\\
&&\!\!\!\!+\bar\uppsi_2^r(p)\gamma^\mu \uppsi_2^s(q)(b_{{\bf p},r}b^\dagger_{{\bf q},s}-b^\dagger_{{\bf q},s}b_{{\bf p},r})e^{i(q-p)\cdot x}\Bigg\}
\nonumber\\
&=&:\,\bar\uppsi\gamma^\mu\uppsi\,:\,-\frac{1}{2(2\pi)^6}\int \frac{d^3{\bf p}}{2E_{\bf p}}p^\mu\sum_{r=1,2}(\bar\uppsi_1^r(p)\uppsi_1^r(p)+\bar\uppsi_2^r(p)\uppsi_2^r(p))\nonumber\\
&=&:\,\bar\uppsi\gamma^\mu\uppsi\,:\eeq
{{ The last equality holds whenever the spinor satisfies  the Dirac equation, using the orthogonality conditions \cite{voja}.

It must be read off the previous expression that
\beq
:\,\bar\uppsi(x)\gamma^\mu\uppsi(x)\,:&=&\frac12[\bar\uppsi(x),\gamma^\mu\uppsi(x)]=\frac12[\bar\uppsi_\alpha(x),(\gamma^\mu)_{\alpha\beta}\uppsi_\beta(x)].\label{reado}\eeq
This quantity can be identified with a current, \beq
J^\mu(x)=\; :\bar\uppsi(x)\gamma^\mu\uppsi(x): \;.\eeq
Another way to see it follows from the fact that the fields, as quantum mechanical operators, satisfy anti-commutation relations to reflect the Fermi-Dirac statistics that the underlying particles obey. In this case, the appropriate operator ordering for the current is antisymmetrization  \cite{das}. 
The action of the charge conjugation operator $C$ reads
\beq
\uppsi(p)\overset{C}{\mapsto}\uppsi^C(p)&=&\eta_\uppsi C\bar\uppsi^\intercal,\qquad\qquad
\bar\uppsi(p)\overset{C}{\mapsto}\bar\uppsi^C(p)=-\eta_\uppsi^* \bar\uppsi^\intercal C^{-1},\eeq
with $|\eta_\uppsi|^2=1, C^\dagger C={\rm id}$ and $C^\intercal = -C$. 
Thus, the normal order current reads 
\beq\label{2ndcurrent}
J^\mu(x)&=&:\,\bar\uppsi(x)\gamma^\mu\uppsi(x)\,:=\frac12\bar\uppsi(x)\gamma^\mu\uppsi(x)-\uppsi^\intercal(x)(\gamma^\mu)^\intercal\bar\uppsi^\intercal(x).\eeq

For computing the scalar bilinear covariant,  the usual definitions are taken into account,
\beq
 \uppsi_\alpha(x)= \uppsi_\alpha^+(x)+\uppsi_\alpha^-(x), \qquad\qquad\bar\uppsi_\alpha(x)= \bar\uppsi_\alpha^+(x)+\bar\uppsi_\alpha^-(x),
\eeq
where
\begin{subequations}
\beq\label{1psigen1}
 \uppsi_\alpha^+(x)&=&\int \frac{d^3{\bf p}}{(2\pi)^3\sqrt{2E_{\bf p}}}\sum_{s=1,2}\;a_{{\bf p},s} \uppsi_{1\alpha}^s(p)\,e^{-ip\cdot{x}},\\
 \uppsi_\alpha^-(x)&=& \int \frac{d^3{\bf p}}{(2\pi)^3\sqrt{2E_{\bf p}}}\sum_{s=1,2}\;b^\dagger_{{\bf p},s} \uppsi_{2\alpha}^s(p)\,e^{i{p}\cdot{x}}\label{psimais}\\
 \bar\uppsi_\alpha^+(x)&=&\int \frac{d^3{\bf p}}{(2\pi)^3\sqrt{2E_{\bf p}}}\sum_{s=1,2}\;b_{{\bf p},s} \bar\uppsi_{2\alpha}^s(p)\,e^{-ip\cdot{x}},\label{psimenos}\\
 \bar\uppsi_\alpha^-(x)&=& \int \frac{d^3{\bf p}}{(2\pi)^3\sqrt{2E_{\bf p}}}\sum_{s=1,2}\;a^\dagger_{{\bf p},s} \bar\uppsi_{1\alpha}^s(p)\,e^{i{p}\cdot{x}}. \label{2psidual}
 \eeq
 \end{subequations}
Hence, 
\beq
:\, \uppsi_\alpha^+(x)\bar\uppsi_\beta^+(y)\,:\;&=&\;:\,(\uppsi_\alpha^+(x)+\uppsi_\alpha^-(x))(\bar\uppsi_\beta^+(y)+\bar\uppsi_\beta^-(y))\,:\;
%&=&\;:\,\left(\uppsi_\alpha^+(x)\bar\uppsi_\beta^+(y)+\uppsi_\alpha^+(x)\bar\uppsi_\beta^-(y)+\uppsi_\alpha^-(x)\bar\uppsi_\beta^+(y)+\uppsi_\alpha^-(x)\bar\uppsi_\beta^+(y)\right)\,:\nonumber\\
%&=&\left(\uppsi_\alpha^+(x)\bar\uppsi_\beta^+(y)-\uppsi_\beta^-(y)\uppsi_\alpha^+(x)+\bar+\uppsi_\alpha^-(x)\bar\uppsi_\beta^+(y)+\uppsi_\alpha^-(x)\bar\uppsi_\beta^+(y)\right)\nonumber\\
%&=&\uppsi_\alpha(x)\bar\uppsi_\beta(y)-\left[\uppsi_\alpha^+(x), \bar\uppsi_\beta^-(y)\right]_+\nonumber\\
=\uppsi_\alpha(x)\bar\uppsi_\beta(y)+iS_{\alpha\beta}^+(x-y),
\eeq
where the propagator reads $
\frac12\int\frac{d^3p}{(2\pi)^3}\frac{(\slash{\!\!\!p}\pm m)_{\alpha\beta}}{2m} e^{-ip\cdot(x-y)}=-iS^\pm_{\alpha\beta}(x-y).$ 
Analogously, 
\beq\label{2ndpsipsi}
:\, \bar\uppsi_\beta(y)\uppsi_\alpha(x)\,:\;&=&\bar\uppsi_\beta(y)\uppsi_\alpha(x)+iS^-_{\alpha\beta}(x-y).
\eeq

Eqs. (\ref{reado}, \ref{2ndpsipsi}) are the second quantized bilinears that are sufficient to construct 3-level 
diagrams, involving terms 
\beq
:\bar\uppsi\uppsi:\,,\,:\bar\uppsi(x)\gamma^\mu\uppsi(x'):\,,
:\bar\uppsi\uppsi\phi:\,,:\bar\uppsi\slash{\!\!\!A}\uppsi:\,, :\bar\uppsi\slash{\!\!\!\omega}\uppsi:\,,
\eeq
for $\slash{\!\!\!A}$ and $\slash{\!\!\!\omega}$ being 
the electromagnetic potential and the spin connection, respectively. 

Now, the bilinear covariant $S^{\mu\nu}$ reads, 
\beq\label{nspin}
:\,\bar\uppsi(x)\gamma^{\mu\nu}\uppsi(x)\,:
&\!=\!&\!\frac{1}{(2\pi)^6}\int\frac{d^3{\bf p}d^3{\bf q}}{\sqrt{2E_{\bf p}2E_{\bf q}}}\!\!\sum_{r,s=1,2}\Bigg\{
\bar\uppsi_1^r(p)\gamma^{\mu\nu} \uppsi_1^s(q)a_{{\bf q},s}a^\dagger_{{\bf p},r}e^{i(p-q)\cdot x}\nonumber\\
+\bar\uppsi_1^r(p)\gamma^{\mu\nu}&&\!\!\!\!\!\!\!\!\!\! \uppsi_2^s(q)a^\dagger_{{\bf p},r}b^\dagger_{{\bf q},s}e^{i(p+q)\cdot x}\!+\!\bar\uppsi_2^r(p)\gamma^{\mu\nu} \uppsi_1^s(q)b_{{\bf p},r}a_{{\bf q},s}e^{-i(p+q)\cdot x}\!+\!\bar\uppsi_2^r(p)\gamma^{\mu\nu} \uppsi_2^s(q)b_{{\bf p},r}b^\dagger_{{\bf q},s}e^{i(q-p)\cdot x}\Bigg\}. 
\eeq
It is a general expression, however it contains all the ingredients for the classification of the quantum fields. 
The second quantized classification of spinors resembles the first quantized one, however it is more strict. In fact, one can straightforwardly realize that for spinors in the same Lounesto spinor class, two terms $\bar\uppsi_1^r(p)\gamma^\mu\gamma^\nu \uppsi_1^s(q)$ and $\bar\uppsi_2^r(p)\gamma^\mu\gamma^\nu \uppsi_2^s(q)$ of Eq. (\ref{nspin}) disappear, if for instance they are dipole spinors.
% However, even assuming that the spinors $\uppsi_1^r(p)$ and $\uppsi_2^s(q)$ are also dipoles, 
%in order that $\uppsi(x)$ be a dipole, we must impose that 
%\beq
%\bar\uppsi_1^r(p)\uppsi_2^s(q)=0=\bar\uppsi_2^r(p) \uppsi_1^s(q).\label{spin00}\eeq
%One must assume that there is no mixture. 
%Clearly, if the spinor field dynamics is taken into account, Eq. (\ref{spin00}) can be satisfied, for instance, for the very particular case when 
%$\uppsi_1^r(p)=u_r(p)$ and $\uppsi_2^s(q)=v_s(p)$ are the textbook standard Dirac spinors. 

Going further, the second quantized pseudovector  is given by
\beq\label{npseudo}
:\,\bar\uppsi(x)\gamma^\mu\gamma^{0123}\uppsi(x)\,:
&\!=\!&\!\frac{1}{(2\pi)^6}\!\!\int\!\!\! \frac{d^3{\bf p}d^3{\bf q}}{\sqrt{2E_{\bf p}2E_{\bf q}}}\sum_{r,s=1,2}\!\Bigg\{
\bar\uppsi_1^r(p)\gamma^\mu\gamma^{0123} \uppsi_1^s(q)a_{{\bf q},s}a^\dagger_{{\bf p},r}e^{i(p-q)\cdot x}\nonumber\\&&\!\!\!\!\!\!\!\!\!\!+\bar\uppsi_1^r(p)\gamma^\mu\gamma^{0123} \uppsi_2^s(q)a^\dagger_{{\bf p},r}b^\dagger_{{\bf q},s}e^{i(p+q)\cdot x}+\bar\uppsi_2^r(p)\gamma^\mu\gamma^{0123} \uppsi_1^s(q)b_{{\bf p},r}a_{{\bf q},s}e^{-i(p+q)\cdot x}\nonumber\\&&\!\!\!\!\!\!\!\!\!\!+\bar\uppsi_2^r(p)\gamma^\mu\gamma^{0123} \uppsi_2^s(q)b_{{\bf p},r}b^\dagger_{{\bf q},s}e^{i(q-p)\cdot x}\Bigg\}.
\eeq
%whereas 
%the pseudoscalar, fully analysed in Sect. III.B reads
%\beq\label{npseudos}
%\!\!\!\!:\,\bar\uppsi(x)\gamma^{0123}\uppsi(x)\,:
%&\!=\!&\!\frac{1}{(2\pi)^6}\!\!\int\!\!\! \frac{d^3{\bf p}d^3{\bf q}}{\sqrt{2E_{\bf p}2E_{\bf q}}}\!\!\sum_{r,s=1,2}\!\Bigg\{
%\bar\uppsi_1^r(p)\gamma^{0123} \uppsi_1^s(q)a_{{\bf q},s}a^\dagger_{{\bf p},r}e^{i(p-q)\cdot x}+\bar\uppsi_1^r(p)\gamma^{0123} \uppsi_2^s(q)a^\dagger_{{\bf p},r}b^\dagger_{{\bf q},s}e^{i(p+q)\cdot x}\nonumber\\
%&&\!\!\!\!+\bar\uppsi_2^r(p)\gamma^{0123} \uppsi_1^s(q)b_{{\bf p},r}a_{{\bf q},s}e^{-i(p+q)\cdot x}+\bar\uppsi_2^r(p)\gamma^\mu\gamma^{0123} \uppsi_2^s(q)b_{{\bf p},r}b^\dagger_{{\bf q},s}e^{i(q-p)\cdot x}\Bigg\}.
%\eeq
The remain bilinears may be computed and investigated as in the previous sections. All of them, considered together, may be used as a starting point in the derivation of a quantum reconstruction theorem.  

\section{Conclusions}

\textcolor{black}{In this paper we have started to analyse the problem of spinor classification in a second quantized framework. Our approach is perturbative and the main object of study are the scalar and pseudoscalar bilinears, together with the Feynman propagator. Within these limits}
the second quantized paradigm has been extended, in order to encompass all the spinors in the Lounesto classification. It encodes \textcolor{black}{in}  particular the well known cases of the Dirac, Weyl, and Majorana quantum fields. The second quantized quantum  field uses general Lounesto's spinors in the expansion of the quantum field, with the only assumption that the arbitrary spinors are assumed to satisfy the Klein-Gordon equation. Nevertheless, no linear first-order equation 
is assumed \emph{a priori}. 
Once established the arbitrary  quantum field expansion in Eq. (\ref{psigen}), we analyzed the subsequent possibilities throughout the paper, to define the classes of regular and singular quantum spinor fields in the second quantized paradigm. If one insists the spinors to be solutions of the Dirac equation in momentum space, for example, such an equation, for plane waves, may lead to a vanishing pseudoscalar, so further simplification might occur in the classification provided in Sect. IV.

The key  \textcolor{black}{ingredient} for deriving the second quantized quantum  field is the reconstruction  \textcolor{black}{theorem}, that makes the analysis 
of the spinors, in the quantum field expansion, to rely 
on the bilinear covariants (and then on the Lounesto's spinor classes themselves). In fact, the reconstruction theorem plays a prominent role in the refinement of all the ramifications that arise in the intricate analysis. Delving into a precise formulation of a  quantum spinor field classification,  \textcolor{black}{which leads} to a further question of classifying the spinors in momentum space, we  \textcolor{black}{have} classified the quantum fields into singular and regular second quantized fields.  Hence, the usual plane wave expansion of a quantum field for a free field  \textcolor{black}{have been} extended, to comprise all the regular and singular spinors in the first quantized formalism. Our conclusions point to a richer classification of  second quantized regular and singular quantum fields. Indeed, we observe that subclasses of regular and singular quantum fields can be split into more subclasses of fields. Thereafter, the calculations of $n$-point functions and propagators  \textcolor{black}{have been} extended to any  spinor in the Lounesto's classification that satisfies the Dirac equation. It is worth to mention that such kind 
of spinors has representatives in the regular and singular spinors in the Lounesto's classification, having examples in all classes of regular spinors \cite{daSilva:2012wp}, and also there are  flag-dipole type-4 singular spinors \cite{esk} and flagpole type-5 singular spinors
\cite{daRocha:2016bil} that satisfy the Dirac equation.
Furthermore,  the $n$-point functions and propagators  \textcolor{black}{have been} derived for arbitrary eigenspinors of the charge conjugation operator that have mass dimension 3/2, in the four-dimensional Minkowski spacetime. In addition, explicit expressions for the normal ordered  bilinear covariants of arbitrary quantum fields  \textcolor{black}{have been} obtained, providing further possibilities towards second quantized version of the reconstruction theorem. 
At this moment it is \textcolor{black}{perhaps premature}  to further split the singular and regular classes of second quantized spinor fields, since Sect. VI was already devoted to construct and derive
the normal ordered bilinear covariants that can encompass tree-level diagrams, that shall be useful to, eventually, construct Lagrangians for arbitrary quantum fields. Possibly refining the found classes is not the most effective way, since we can define operators that send one Lounesto's class to another  \cite{Cavalcanti:2014uta}.

Once established the classification of second quantized spinor fields into regular and singular classes  
in the four-dimensional Minkowski space, an analogous second quantized  classification on 
other types of spaces can be further considered, including the new spinor classes on compactifications AdS$_4\times M^7$, recently found by Ref.  \cite{bonora,Bonora:2015ppa}.

\subsection*{Acknowledgements}
JMHS thanks to CNPq (304629/2015-4; 445385/2014-6) for partial financial support.
RdR~is grateful to SISSA for the hospitality and to INFN, to CNPq (Grant No. 303293/2015-2), and to FAPESP (Grant No. 2017/18897-8), for partial financial support.

\end{document}